# Effects of Varying Incident Wave Inclination and Azimuthal Angles on Multi-Dimensional Ground Response Analyses at the Delaney Park Downhole Array Site

Nishkarsha Dawadi[*] and Brady R. Cox

Department of Civil and Environmental Engineering, Utah State University, Logan, Utah, USA

**ABSTRACT**

Even when large-scale, site-specific, three-dimensional (3D) subsurface models are used to represent spatial variability, multi-dimensional numerical ground response analyses (GRAs) at downhole array sites continue to exhibit significant amplitude discrepancies between simulated theoretical transfer functions (TTFs) and recorded empirical transfer functions (ETFs). At the Delaney Park Downhole Array (DPDA) site, ETFs computed from small-strain ground motions show significantly lower amplitudes at the fundamental frequency ($f_0$) than predicted by TTFs, suggesting greater apparent attenuation from wave scattering and destructive interference than is currently captured in multi-dimensional GRAs. However, most prior two-dimensional (2D) and 3D GRAs assume vertically propagating, horizontally polarized shear-wave input, neglecting scenarios where seismic waves approach at inclined angles and from different azimuths. This study evaluates the effects of varying inclination and azimuthal input angles in 2D and 3D GRAs at DPDA to assess whether non-vertical and directional wave incidence increase apparent damping and improve agreement with observed ETFs. Two approaches for modeling inclined waves, the Input Lag Method (ILM) and the Inclined Domain Method (IDM), are first compared, and the ILM is shown to be more effective and computationally efficient for large-scale models. Using ILM, a parametric study is conducted across a range of inclination and azimuthal angles. Results indicate that inclination angles up to 15° produce only minor reductions in TTF amplitudes near $f_0$, with some cases yielding slightly improved ETF agreement. Larger inclination angles significantly reduce amplitudes near $f_0$, but also cause systematic shifts in $f_0$ to higher frequencies that are not observed in the ETFs. These findings indicate that moderately inclined waves do not significantly reduce small-strain TTF amplitudes, and that steeply inclined incidence is unlikely to represent the ground motions recorded at DPDA. Azimuthal input variation in 3D GRAs produces relatively minor changes in TTFs, primarily affecting trough amplitudes while leaving $f_0$ and higher-mode peaks largely unchanged.

## 1. Introduction

Numerical ground response analyses (GRAs) are used to evaluate how seismic waves are altered in frequency content and amplitude as they travel from bedrock to the surface. The degree of seismic wave alteration is controlled by site-specific material properties (e.g., modulus and damping) as well as subsurface layering and wave-scattering effects. Although one-, two-, and three-dimensional (1D, 2D, and 3D) GRAs exist, most engineering practice still relies on simplified 1D GRAs that assume horizontal, laterally-uniform layers and vertically-propagating, horizontally-polarized shear ($S_{VH}$) waves as input. Recent studies have shown that 1D GRAs frequently fail to reproduce observed site response at many borehole array sites (e.g., Hallal et al., 2022b; Tao and Rathje, 2020; Afshari and Stewart, 2019; Pilz and Cotton, 2019; Kim and Hashash, 2013; Kaklamanos et al., 2013; Thompson et al., 2012; Thompson et al., 2009), typically overpredicting amplification near the fundamental and first few higher-mode peaks. This bias has largely been attributed to the inability of 1D modeling approaches to represent real subsurface heterogeneity that would lead to greater wave-scattering and apparent attenuation if modeled properly.

---

[*] Corresponding author:
Email: nishkarsha.dawadi@usu.edu (N. Dawadi)

Seismic wave attenuation arises from intrinsic, geometric, and apparent damping (Zywicki 1999). Intrinsic damping, which can be measured in laboratory tests, reflects energy dissipation within soils, while geometric damping describes amplitude decay as waves spread outward from the source. Apparent damping results from wave-scattering processes caused by subsurface heterogeneity (Rix et al., 2000; Spencer et al., 1977; O'Doherty and Anstey, 1971) and is strongly site-specific, often dominating the in-situ small-strain intrinsic damping estimates (Abbas et al. 2025). To approximate the energy loss associated with apparent damping, many researchers have tried to artificially inflate the intrinsic small-strain damping ratio ($D_{min}$) used in 1D GRAs. Proposed approaches include raising $D_{min}$ to 2–5% (Cabas et al., 2017; Tsai and Hashash, 2009), adding fixed $D_{min}$ increments of 1–4% (Yee et al. 2013), applying $D_{min}$ multipliers of 2–6 times (Tao and Rathje, 2019; Zalachoris and Rathje, 2015), and using kappa-based $D_{min}$ formulations (Afshari and Stewart 2019). While these methods often reduce overamplification at the fundamental frequency ($f_0$) peak, they commonly overdamp higher-mode peaks, resulting in underpredicting high-frequency amplification (Hallal et al., 2022b; Kaklamanos et al., 2020). Other efforts to incorporate spatial variability and its associated apparent attenuation effects within 1D frameworks, such as shear wave velocity ($V_s$) randomization (Toro, 2022; 1995), randomizing cumulative shear-wave travel time (Hallal et al., 2022a; Passeri et al., 2020), and using multiple $V_s$ profiles from surface wave testing (Teague et al., 2018), have also often resulted in overdamping at the $f_0$ and/or higher-mode peaks, sometimes excessively (Chang et al., 2022; Hallal et al., 2022b; Hallal and Cox, 2021a; Teague et al., 2018). A few studies have combined $V_s$ randomization with damping adjustments and found that multiple approaches to account for spatial variability interact additively, requiring each to be used at reduced levels (Rathje et al., 2010; Rodriguez-Marek et al., 2017). Stochastic random-field approaches have also been used to expand 1D profiles into 2D or 3D models, and while these approaches can be effective at reducing overamplification by broadening the $f_0$ peak, they generally yield models that are inconsistent with larger-scale structural features such as dipping layers or laterally discontinuous stratigraphy (de la Torre et al., 2022; Hu et al., 2021; Thompson et al., 2009).

More recently, large-scale, site-specific, pseudo-3D subsurface $V_s$ models built using the horizontal-to-vertical spectral ratio (H/V) geostatistical approach (Hallal and Cox, 2021a; 2021b) have been used to perform 2D and 3D GRAs (Dawadi et al., 2024; Hallal and Cox, 2023). These studies have demonstrated that incorporating large lateral extents (hundreds of meters to over 1 km) into multi-dimensional GRAs is important at many sites and can improve agreement between theoretical transfer functions (TTFs) and empirical transfer functions (ETFs) calculated from small-strain ground motions recorded at downhole arrays. This improvement occurs as a result of lowering amplification and broadening resonant peaks due to wave scattering off major dipping impedance contrasts that are well captured by the pseudo-3D $V_s$ models developed using the H/V geostatistical approach. However, pseudo-3D $V_s$ models often rely on a single downhole $V_s$ profile that is stretched and compressed to match spatial variability in resonant frequencies across the site and, therefore, may miss other smaller/localized heterogeneities that contribute to apparent damping. As such, if the site is highly variable, the pseudo-3D $V_s$ approach may not fully capture important wave scattering features. Dawadi et al. (2026b) addressed this limitation by combining small-strain $D_{min}$ multipliers with a frequency-dependent Rayleigh mass-only damping formulation when performing 2D GRAs, resulting in substantially improved agreement between simulated TTFs and recorded ETFs at the Delaney Park Downhole Array (DPDA) site. Dawadi et al. (2026c) extended this approach to three additional downhole array sites and found that similar improvements were consistently observed. While this approach of artificially inflating small-strain $D_{min}$ and using a frequency-dependent damping formulation seems to work well from a practical application perspective, it still does not properly address some of the unmodeled physics of wave propagation in a heterogeneous media that could potentially be contributing to wave scattering and apparent attenuation. For example, Dawadi et al. (2026c) reported that $D_{min}$ multipliers were used to not only account for wave scattering and apparent attenuation from unmodeled spatial variability, but also the potential effects of inclined, non-vertical wave incidence that are not typically modeled in GRAs.

An important factor that can influence attenuation and overall site response modeling is the manner in which seismic waves enter the numerical domain. Most previous 1D, 2D, and 3D GRA studies have applied ground motions consistent with vertically-propagating, horizontally-polarized shear waves that are input uniformly across the model base. Although spatial variations in bedrock geometry can still generate non-vertical wave incidence within the model (Dawadi et al., 2026c; Dawadi et al., 2024; Hallal and Cox, 2023), it remains likely that GRA predictions would differ if the input motion itself were prescribed as non-vertical rather than purely vertical. Only a limited number of studies have directly investigated inclined or non-vertically incident waves within multi-dimensional GRAs performed at borehole array sites. For example, Dawadi et al. (2026a) performed 3D GRAs at the I-15 Downhole Array (I15DA) using the epicentral locations of recorded events to constrain input azimuth and systematically evaluating the effects of varying incidence angles on site response. Their results showed that non-vertical wave incidence can produce measurable changes in both peak amplification and peak frequencies, but predominantly at larger incidence angles. Similarly, Eskandarighadi et al. (2026) incorporated non-vertical incidence and spatially variable ground motions into 2D simulations at the Treasure Island Downhole Array (TIDA), demonstrating that inclined waves can alter TTF amplitudes and modal characteristics, although the resulting improvements in mean predictions were modest for that site.

In this study, we perform 2D and 3D GRAs by directly modeling input wave incidence angle in a parametric study to investigate the impact on site response predictions at the DPDA site, which is one of two downhole array sites we have studied where ETF amplitudes from small-strain recorded ground motions are particularly low and difficult to match. We begin by introducing and comparing two approaches for applying inclined waves, the Input Lag Method (ILM) and the Inclined Domain Method (IDM). Based on this evaluation, we adopt the more effective ILM method to conduct 2D GRAs along the most heterogeneous cross-section extracted from the DPDA pseudo-3D $V_s$ model for a wide range of wave inclination angles. We then extend the approach to 3D GRAs using the full pseudo-3D $V_s$ model, exploring how varying inclination and wave arrival azimuthal angles influence modal behavior, amplification patterns, and overall agreement with the recorded ETFs at the DPDA site.

## 2. Site Description and Empirical Transfer Functions (ETFs)

The DPDA site is located in downtown Anchorage, Alaska, where the subsurface consists of 10–15 m of glacial outwash overlying approximately 30 m of Bootlegger Cove Clay and underlain by glacial till (Combellick 1999). The Bootlegger Cove Clay includes interbedded cohesive and non-cohesive facies, with the former exhibiting high sensitivity and the latter being susceptible to liquefaction (Badal et al. 2004). Strong-motion sensors are installed at depths of 0, 4.6, 10.7, 18.3, 30.5, 45.4, and 61 m, though only the 0- and 61-m sensors were used in this study. The 1D $V_s$ profile, developed from downhole seismic testing by Thornley et al. (2019), served as the reference profile for constructing a pseudo-3D $V_s$ model. This profile is shown along with the soil stratigraphy of the DPDA in Figure 1a.

The DPDA pseudo-3D $V_s$ model was originally developed by Hallal and Cox (2021a) using the H/V geostatistical approach, which integrates a single measured $V_s$ profile with a dense grid of H/V spectral-ratio measurements. This model was subsequently refined by Dawadi et al. (2024) using some additional H/V measurements. For constructing the present version of the DPDA pseudo-3D $V_s$ model (Figure 1b), 108 H/V measurements were used to generate a domain measuring 1.6 × 1.6 km laterally and 80 m in depth. The H/V measurement time records are publicly available in the DesignSafe data repository (Hallal et al., 2025). This model consists of 25.6 million 2-m cubic elements, with $V_s$ values ranging from ~200 m/s near the surface to ~1450 m/s at the clay–till boundary. It can be seen in Figure 1b that the southeastern portion of the model contains shallower glacial till, reflected by the higher H/V-measurement-based fundamental frequencies ($f_{0,H/V}$), which is consistent with the stratigraphy described by Combellick (1999). The most heterogeneous cross-section, oriented at the azimuthal direction Az = 165° (Figure 1c), was used for conducting 2D GRAs in this study. For more information on the statistical approach used for extracting the most heterogeneous cross-section from a pseudo-3D $V_s$ model, readers are referred to Dawadi et al. (2026c).

Performing 3D GRAs using the 25.6-million-element model is computationally demanding; therefore, the pseudo-3D model was re-discretized to 5-m cubic elements, resulting in approximately 1.64 million elements. While this increase in element size substantially lowers computational demands, it also decreases the model's maximum resolvable frequency. Following the guidance of Lysmer & Kuhlemeyer (1969), the maximum element size for 2D or 3D wave-propagation analyses should not exceed one-eighth of the wavelength of the slowest body wave. Considering that our simulations involve substantial wave scattering, we adopt a more conservative requirement of at least 10 elements per wavelength to adequately capture obliquely-incident waves. Based on this criterion, the maximum resolvable frequencies for the 2-m and 5-m grid models are 11.2 Hz and 4.48 Hz, respectively. These frequency limits are hereafter denoted as $f_{s,2m}$ and $f_{s,5m}$, respectively, and are indicated in the ETF plots presented later.

ETFs for DPDA were computed using 56 low-amplitude recordings (28 events × 2 horizontal components) with peak ground accelerations of 0.001–0.01 g, local magnitudes between 3.0 and 5.1, and source-to-site distances ranging from 10 to 96 km. The event locations are distributed quite equally in terms of azimuth around the DPDA site, as shown in Figure 2a. ETFs were defined as the ratio of the Fourier amplitude spectrum (FAS) of the recorded horizontal acceleration at the surface to that at the deepest downhole sensor (61 m; refer to Figure 1a). All FAS were smoothed using the Konno & Ohmachi (1998) window with bandwidth, $b = 75$ to reduce noise and improve interpretability. Because ETFs are derived from two horizontal components for each event, several approaches can be used to combine these data for evaluating

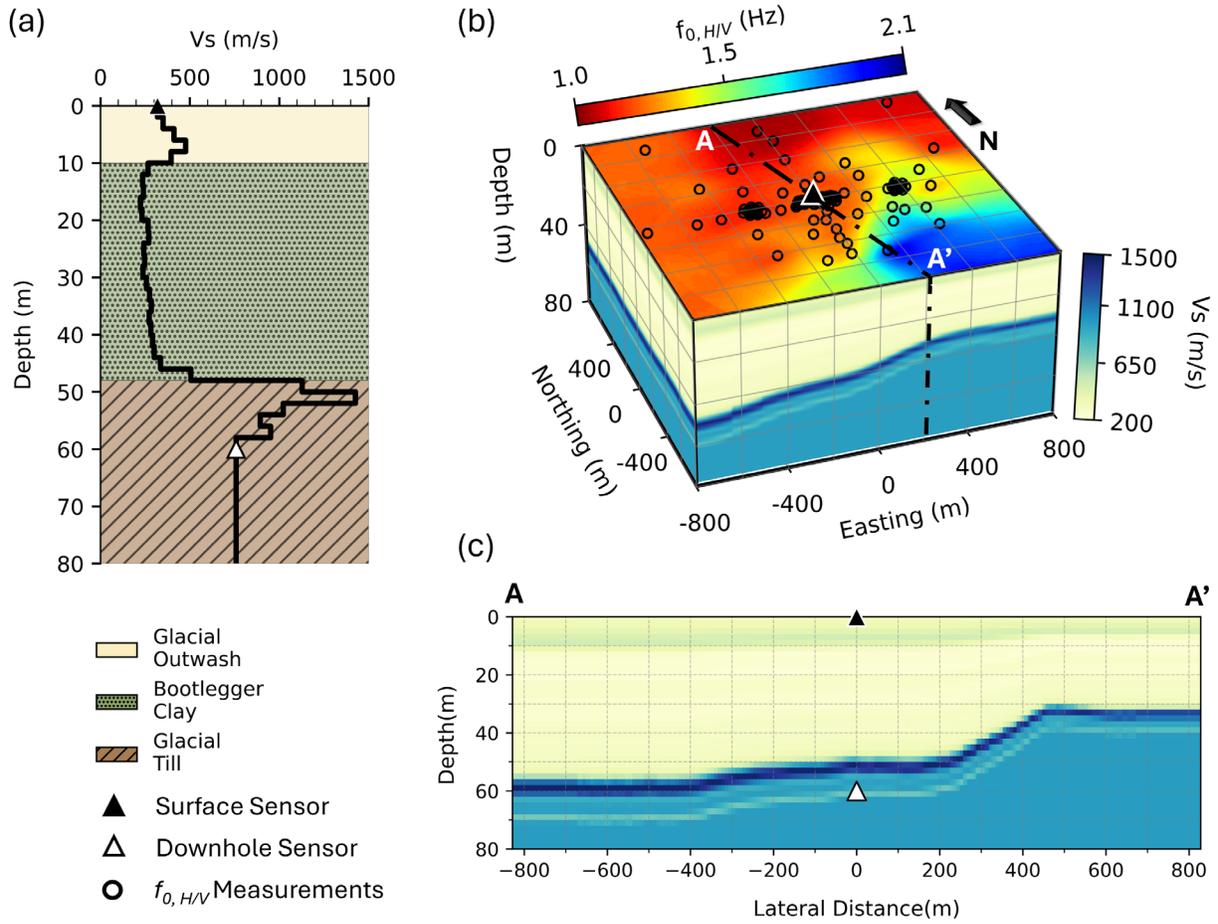

**Figure 1.** (a) 1D Vs profile from Thornley et al. (2019) and soil layers at the DPDA site; (b) pseudo-3D Vs model of the DPDA site developed using the H/V geostatistical approach; and (c) the A-A' cross-section extracted from the pseudo-3D Vs model along the azimuthal direction Az = 165°.

GRA results. To examine the sensitivity to the chosen combination method, multiple ETF representations were evaluated. First, lognormal median ETFs were computed separately for the EW and NS components, $LM_{ETF,EW}$ and $LM_{ETF,NS}$, respectively, by taking the median across all individual ETFs for each direction. The associated natural logarithmic standard deviations, $\sigma_{\ln ETF,EW}$, and $\sigma_{\ln ETF,NS}$, respectively, were calculated at each frequency to quantify variability. These component-specific median ETFs, along with their ±1 σ bounds, are shown in Figures 2b and 2c, respectively. Next, an overall lognormal median ETF, $LM_{ETF,EW\&NS}$, was computed by combining all individual ETFs from both horizontal components, together with the associated $\sigma_{\ln ETF,EW\&NS}$, as shown in Figure 2d. Finally, to facilitate evaluation of GRAs conducted along the Az = 165° cross-section, the individual EW and NS component ETFs for each event were rotated into this azimuthal direction and combined to compute a representative lognormal median ETF, $LM_{ETF,Az165}$, along with its corresponding $\sigma_{\ln ETF,Az165}$. These results are shown in Figure 2e.

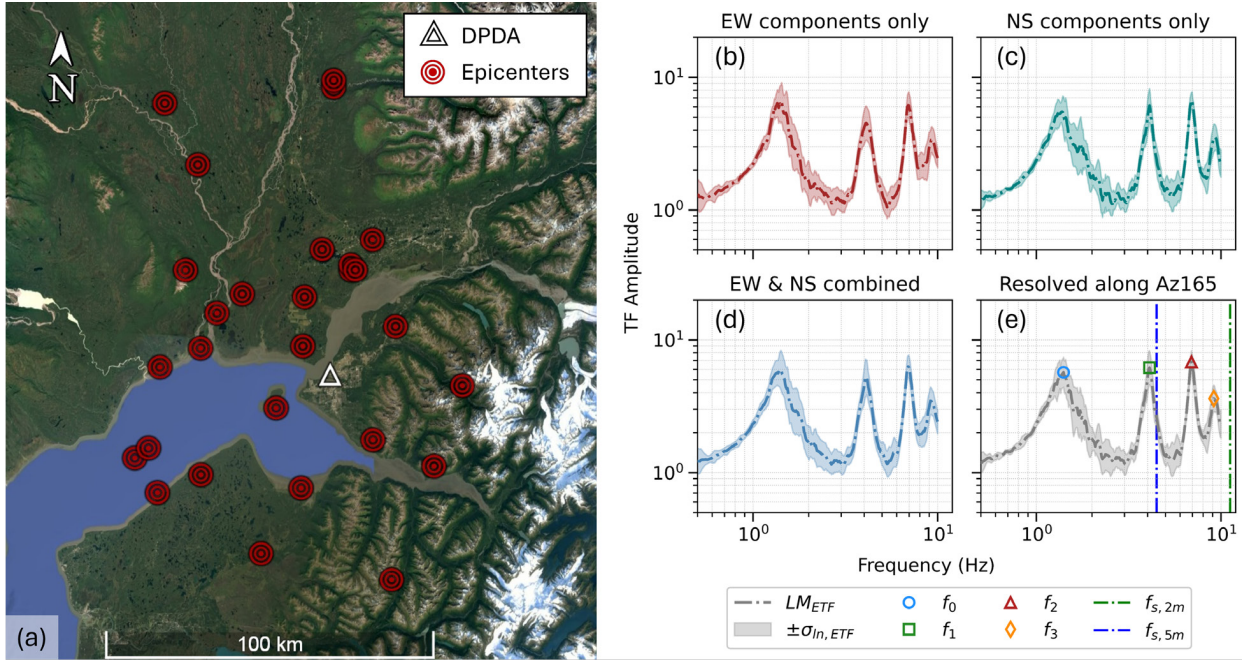

**Figure 2.** (a) Distribution of epicenters of the events used to compute the ETFs in this study. Panels (b) and (c) show the lognormal median ETFs and associated lognormal standard deviation bounds computed using only the EW and NS components of these events, respectively ($LM_{ETF,EW} \pm 1\sigma_{ln\,ETF,EW}$ and $LM_{ETF,NS} \pm 1\sigma_{ln\,ETF,NS}$). Panel (d) shows the corresponding results obtained by combining both horizontal components ($LM_{ETF,EW\&NS} \pm 1\sigma_{ln\,ETF,EW\&NS}$). Panel (e) shows the results obtained after rotating both horizontal components into the Az165° direction ($LM_{ETF,Az165} \pm 1\sigma_{ln\,ETF,Az165}$). Also shown in (e) are the identified resonant frequencies ($f_0, f_1, f_2, f_3$), along with the maximum resolvable frequencies associated with 2-m and 5-m model element discretizations ($f_{s,2m}$ and $f_{s,5m}$).

As evident from Figures 2b–2e, the EW, NS, combined, and Az165° ETFs are similar to each other, not showing significant differences across azimuth. Nonetheless, for consistency with the 2D GRA configuration, $LM_{ETF,Az165}$ and its $\pm 1\sigma_{ln\,ETF,Az165}$ bounds (Figure 2e) are used for model evaluation in this study unless stated otherwise. The fundamental and higher-mode site frequencies ($f_0 = 1.4$ Hz, $f_1 = 4.1$ Hz, $f_2 = 6.9$ Hz, and $f_3 = 9.1$ Hz) observed in the $LM_{ETF,Az165}$ are indicated in Figure 2e. The frequency limits associated with the 2-m and 5-m model element discretization ($f_{s,2m}$ and $f_{s,5m}$) discussed earlier are also shown in Figure 2e as green and blue vertical dash-dot lines, respectively.

## 3. Description of Numerical Model

This study used 2D and 3D linear-viscoelastic simulations performed in FLAC3D version 9.0 (Itasca Consulting Group, 2023), which employs a finite-volume framework. The linear-viscoelastic modeling approach was chosen to isolate the influence of subsurface spatial variability without introducing the additional complexities of modeling nonlinear soil behavior. This strategy aligns with many prior downhole array studies that aim to accurately match small-strain site response before incorporating nonlinear effects (e.g., Hallal and Cox, 2023; Tao and Rathje, 2019). Material properties required for analysis include $V_s$, mass density, Poisson's ratio, and $D_{min}$. A constant mass density of 2000 kg/m³ was assigned to all elements (referred to as zones in FLAC). Poisson's ratio was set to 0.30 in dry soils, 0.48 in saturated soils below the water table (reported at 21 m by Thornley et al., 2019), and 0.30 for rock-like materials with $V_s$ > 760 m/s. $D_{min}$ values for all soil types were calculated using the empirical formulation of Darendeli (2001), considering plasticity index, overconsolidation ratio, and mean effective confining stress. An excitation frequency of 3 Hz was used for computing $D_{min}$. For detailed discussion on the effect of excitation-frequency on $D_{min}$, readers are referred to Dawadi et al. (2026c). The resulting 1D small-strain damping profile at the downhole array was then spatially scaled using the same H/V geostatistical approach applied to the $V_s$ model, thereby producing a pseudo-3D damping model that incorporates spatial variability consistent with the pseudo-3D $V_s$ structure. The DPDA pseudo-3D damping model is described in detail in Dawadi et al. (2026c).

Full Rayleigh damping formulation was implemented, and the damping curves were fit to the target $D_{min}$ values over the frequency range of interest using a root mean square (RMS) minimization procedure. The frequency range of interest extended between the half-amplitude before the ETF $f_0$ peak and the half-amplitude after the $f_3$ peak (approximately 1 – 10 Hz). Quiet (absorbing) boundaries were assigned at the model base, and free-field boundaries were applied along the sides to minimize wave reflections. However, the performance of free-field boundaries is nontrivial when waves approach the model at an inclination, and the associated challenges are discussed in a later section. Because quiet boundaries were used at the base, input motions were applied as stresses. In the present framework, where comparisons are based on transfer functions calculated between the reference depth and the surface, rather than on absolute time histories, it is not necessary to use the exact ground motions recorded by the downhole array as input. Instead, any input motion that contains sufficiently broadband energy across the frequency range of interest is acceptable, and amplitudes are not important, as only linear-viscoelastic properties are modeled prior to calculating TTFs. Following common practice (e.g., Hallal and Cox, 2023), Ricker wavelets were used as the input because they provide a well-controlled broadband signal that is well suited for evaluating frequency-dependent amplification. To ensure adequate energy across the frequency range of interest, a broadband input motion was generated by superimposing five Ricker wavelets with center frequencies logarithmically spaced between 0.5 Hz and 10 Hz (specifically, 0.50, 1.06, 2.24, 4.73, and 10.00 Hz).

## 4. Application of Inclined waves

Before discussing the application of inclined waves, it is important to first briefly review how vertically incident waves are commonly applied at the base of a model for multi-dimensional GRAs. As discussed in an earlier section, an absorbing (or compliant) boundary condition is applied at the model base to attenuate downward-propagating waves reflected from the ground surface. Because these compliant boundaries are implemented using dashpots, seismic energy is introduced into the model in the form of equivalent stresses rather than prescribed velocities. Accordingly, a given particle-velocity time history can be converted to an equivalent input stress using the following expressions:

$$\sigma = -2\,\rho\,V_p\,P_n \qquad (1)$$

$$\tau = -2\,\rho\,V_s\,P_s \qquad (2)$$

where $\sigma$ and $\tau$ are the applied normal and shear stresses, respectively; $\rho$ is the material density; $V_p$ and $V_s$ are the compressional and shear-wave velocities of the medium, respectively; and $P_n$ and $P_s$ are the normal and shear particle-velocity components of the input motion, respectively. These expressions are derived under the assumption of plane-wave propagation. The factor of two arises because, at a compliant boundary, one half of the incident wave energy is absorbed by the dashpots; consequently, the applied stress must be doubled to reproduce the motion that would occur in an infinite medium. In most engineering site response analyses, shear-wave incidence is assumed, while compressional-waves are typically neglected. This assumption is justified because shear waves primarily generate horizontal ground motion, which is responsible for the majority of structural damage during earthquakes. Therefore, all discussions in this study are limited to shear-wave input only.

In contrast to vertically incident waves, inclined waves approach the model base at an oblique angle. Under these conditions, simply applying seismic energy uniformly at the base of the model is no longer appropriate, for reasons discussed later in this section. In principle, the most rigorous approach for simulating inclined waves would be to explicitly model the entire propagation path from the seismic source to the site. However, such an approach is computationally prohibitive for GRAs performed with particular focus on the near-surface using small elements. For example, the 2D cross-section extracted from the pseudo-3D DPDA model is approximately 1.6 km wide and 80 m deep, while the hypocentral depths of the earthquakes considered in this study range from approximately 4 km to 46 km, with epicentral distances ranging from 10 km to 60 km. Explicitly modeling wave propagation over these distances would be computationally prohibitive, even with access to high-performance computing resources. Furthermore, modeling inclined incident waves in site response analyses is inherently challenging because standard boundary condition formulations, particularly those used to simulate infinite lateral extent, are primarily designed for vertical incidence. Introducing inclined waves requires a non-uniform application of input motion at the model base, which produces wavefronts that interact with the boundaries in fundamentally different ways. If not implemented carefully, these interactions can lead to artificial reflections and numerical distortions. To address these challenges, two approaches for simulating inclined incident waves in FLAC3D are evaluated in this study: (1) the Input Lag Method (ILM) and (2) the Inclined Domain Method (IDM). These methods are first discussed in the context of 2D GRAs, followed by a detailed discussion of the preferred approach for full 3D applications.

## 4.1  Input Lag Method (ILM) for 2D GRA

For the ILM, illustrated in Figure 3, the surface of the 2D computational domain remains horizontal, identical to the configuration used for vertically propagating waves. Inclined waves are simulated by applying a chronologically staggered input motion at the model base. To correctly represent an inclined wavefront, the ILM introduces a lag distance that delays the application of the input motion at each base element according to its horizontal position along the direction of wave propagation. This staggered, or ramped, application of the input motion causes the wavefront to sweep across the model base at a prescribed inclination angle ($\theta$), thereby reproducing the physical behavior of an inclined plane wave entering the domain. The time delay at each base element is computed using the lag distance for that element and the $V_s$ of the halfspace at the base of the model.

For an inclined wave approaching the model base (Figure 3) at an angle $\theta$ relative to the vertical, the time delay at the center of the $n^{th}$ base element is given by:

$$\Delta t_n = \frac{\text{Lag Distance}}{V_{s,\text{halfspace}}} = \frac{x_n \sin \theta}{V_{s,\text{halfspace}}} \tag{3}$$

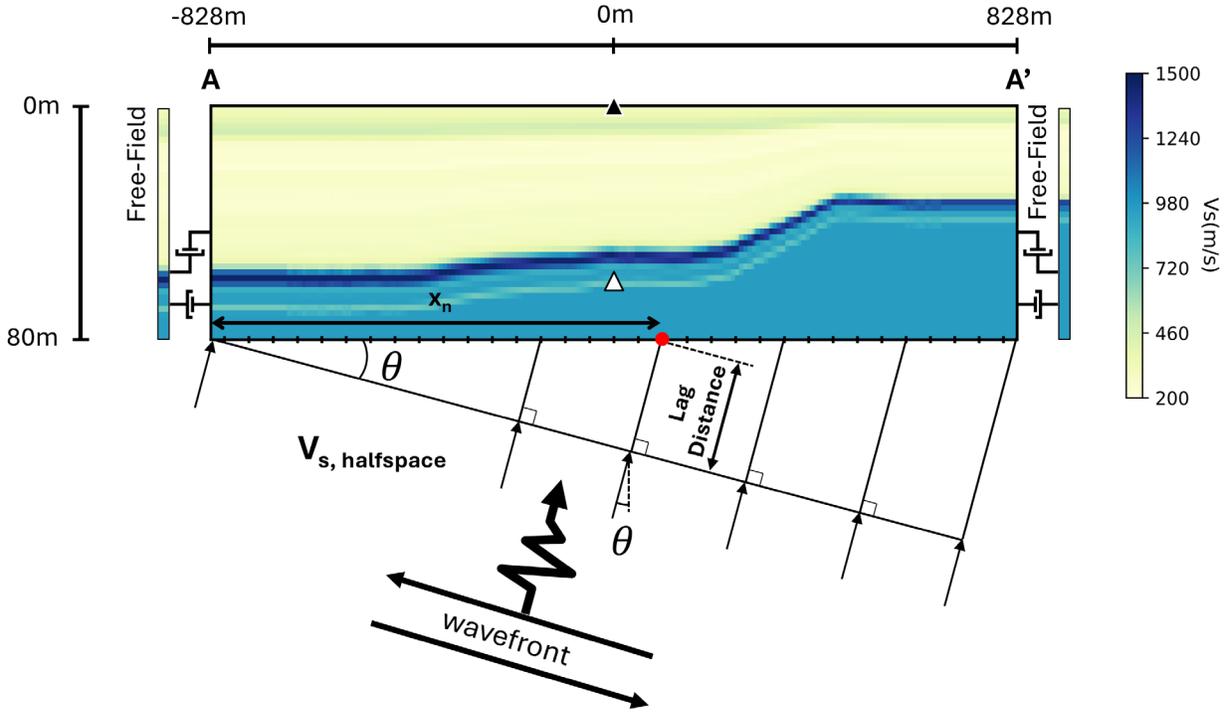

**Figure 3.** Schematic of the ILM used to apply inclined incident waves to the most heterogeneous cross-section (Az165°) at the DPDA site. Free-field boundaries are applied along the sides of the model.

where $x_n$ is the horizontal coordinate of the center of the $n^{th}$ base element measured along the direction of wave propagation, and $V_{s,halfspace}$ is the shear-wave velocity of the halfspace at the base of the model. In addition to the time delay, the incident inclined wave field must be decomposed into horizontal and vertical components consistent with the specified inclination angle $\theta$ prior to staggered application at the model base. This decomposition is performed using the horizontal and vertical scaling factors, $f_x$ and $f_z$, defined in Equation (4):

$$f_x = \cos\theta \quad (4a)$$

$$f_z = -\sin\theta \quad (4b)$$

From Equations (2) and (4), the applied input shear-stress can be written as:

$$\tau_x = \tau \cdot f_x = -2\,\rho\,V_s\,P_s\,\cos\theta \quad (5a)$$

$$\tau_z = \tau \cdot f_z = 2\,\rho\,V_s\,P_s\,\sin\theta \quad (5b)$$

From Equation (5), it is evident that for a vertically propagating incident shear wave ($\theta = 0°$), stress is applied only in the horizontal direction. As the inclination angle increases, a vertical component of particle motion is introduced and increases with increasing inclination angle. The $\tau_x$ and $\tau_z$ components are applied simultaneously at each element but are staggered across the model base at the appropriate time delay, $\Delta t_n$, as defined in Equation (3).

A key consideration when using the ILM is the behavior of the boundary conditions. Although periodic, transmitting, and perfectly matched layer (PML) boundaries offer various advantages for wave propagation problems, each presents limitations when applied to site-specific seismic response modeling. Periodic boundaries require lateral symmetry and continuity across opposing boundaries and are therefore not well suited for spatially heterogeneous domains. Transmitting (viscous) boundaries, such as those based on the Lysmer–Kuhlemeyer formulation, allow outgoing waves to exit the model and reduce artificial reflections; however, they are most effective for waves incident approximately normal to the boundary (Lysmer and Kuhlemeyer, 1969) and may not perform well when waves approach the boundary obliquely. PML boundary formulations provide improved absorption of outgoing waves over a wide range of incidence angles; however, they require additional absorbing layers and careful numerical implementation, which increases model size and computational effort and can be impractical for large-scale models such as the one used in this study (Berenger, 1994; Komatitsch and Tromp, 2003). Furthermore, such formulations are not implemented in commonly used finite-difference codes such as FLAC3D. Conventional free-field boundary formulations simulate an effectively-infinite lateral extent by coupling the side boundaries of the main grid to auxiliary free-field columns through viscous dashpots (Lysmer & Kuhlemeyer, 1969). These free-field columns are designed to reproduce the response of an unbounded medium under 1D wave propagation, typically assuming vertically propagating, horizontally polarized shear waves. In numerical platforms such as FLAC3D, dashpot coefficients and boundary forces are derived from these auxiliary free-field columns. Before conducting analysis on the 2D main grid, a 1D wave is propagated through the 1D free-field columns, and the resulting unbalanced reaction forces are applied to the corresponding boundaries of the main grid. When the main grid is laterally uniform and no surface structures are present, it experiences the same motion as the free-field columns, the dashpots remain inactive, and upward-propagating plane waves pass through the lateral boundaries without distortion. Only when the main-grid motion deviates from the free-field response, such as due to scattering or radiation from surface or subsurface heterogeneities, do the dashpots activate and absorb outgoing energy, thereby limiting artificial wave reflections into the main grid.

In the ILM, however, the incident motion is inclined from the outset, causing wave components to interact with the lateral boundaries at oblique angles. Under these conditions, the free-field response of the main grid is inherently 2D and cannot be represented by boundary formulations derived assuming 1D vertical wave propagation. Because the free-field columns continue to enforce a 1D solution while the main grid is subjected to a 2D inclined input motion, this incompatibility exists from the beginning of the simulation. As a result, the dashpots connecting the free-field columns to the main grid resist the imposed motion, effectively "fighting" the correct input. This nonphysical activation of the lateral boundary dashpots can lead to spurious reflections, artificial distortion of the incident wavefront, and partial trapping of seismic energy along the model. However, as shown below, this theoretical limitation of the free-field boundary conditions is not a significant problem.

### 4.2 Inclined Domain Method (IDM) for 2D GRA

For the IDM (refer to Figure 4), the entire original 2D model domain is rotated by the prescribed incidence angle $\theta$, and the input motion is applied uniformly at the base of an expanded model domain that fully encompasses the original domain as vertically-propagating, horizontally-polarized shear waves. This approach is recommended by Itasca in the FLAC3D documentation for simulating inclined wave incidence. As the vertically propagating waves naturally travel upward through the rotated model, they enter the original model domain along a path that corresponds to an inclined wave with an incidence angle $\theta$ relative to the vertical in the original, unrotated coordinate system.

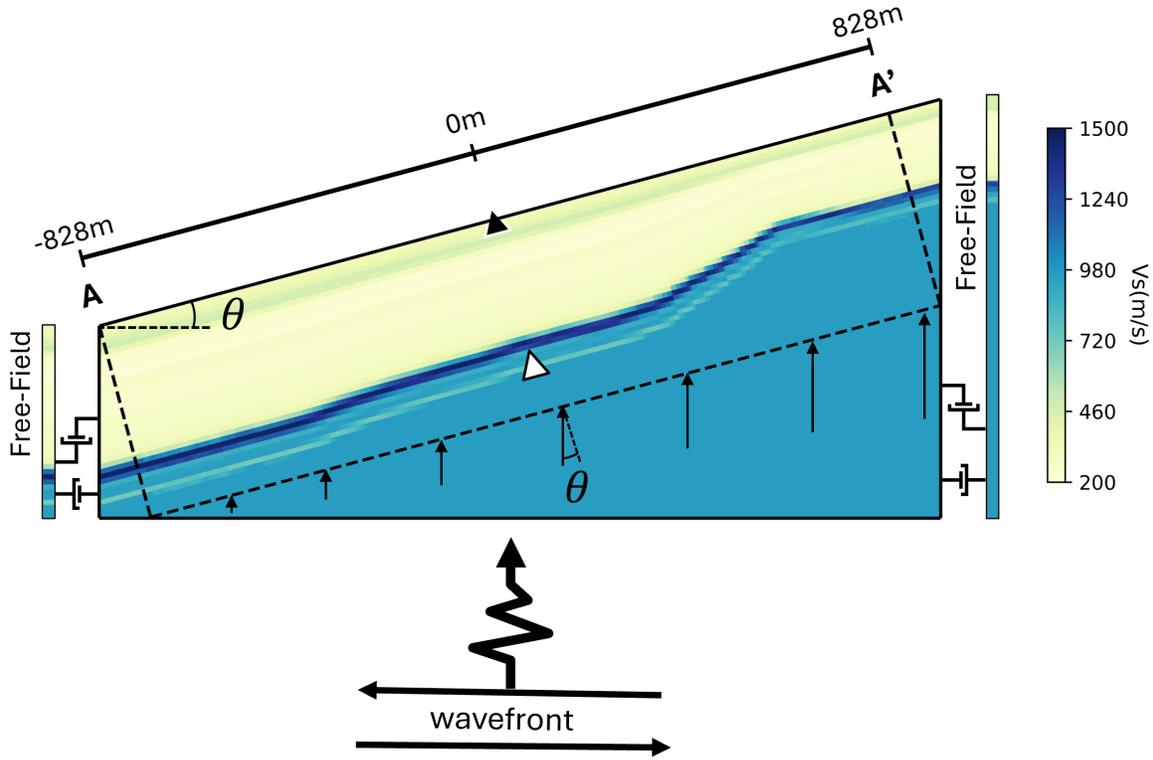

**Figure 4.** Schematic of the IDM used to apply inclined incident waves to the most heterogeneous cross-section (Az165°) at the DPDA site. Dashed lines represent the extent of the original domain, while solid lines indicate the expanded model domain. Free-field boundaries operate normally in the rotated domain.

In this method, the free-field boundaries theoretically remain effective at absorbing outward-propagating waves, at least until the waves hit the sloping ground surface, because the imposed base motion propagates vertically relative to the base of the expanded model domain. However, rotating the model domain requires a substantially larger computational region to fully enclose the rotated geometry. As the rotation angle increases, this expansion leads to a rapid increase in the number of elements. For example, a 30° rotation applied to the Az165° 2D cross-section at DPDA, which extends more than 1.6 km laterally and 80 m in depth, increases the total element count by more than 500%, resulting in a substantial increase in computational cost. In addition, rotating the domain can introduce significant mesh distortion near the model boundaries. To maintain acceptable element aspect ratios and ensure smooth mesh transitions, the computational region must be further enlarged so that the rotated model is embedded within a well-refined mesh. Furthermore, once the wavefront reaches the free surface of the rotated domain, complex scattering paths can introduce reflections that are also difficult to control with the free-field boundaries.

### 4.3 Comparison of ILM and IDM for 2D GRA

To evaluate the performance of the two methods, both ILM and IDM were applied to a laterally-homogeneous 2D cross-section constructed from the 1D DPDA $V_s$ profile (refer to Figure 5). This controlled setting ensures that any observed boundary reflections arise solely from numerical and boundary-condition artifacts, rather than from laterally-variable stratigraphic complexities. Initial simulations were performed for very small inclination angles (< 5°) to ensure that both approaches yielded virtually identical results. While not shown here, these simulations validated the similarity and consistency between the ILM and IDM modeling approaches for small inclination angles. Additional simulations were then performed for three larger inclination angles: 5°, 15°, and 30°. For each case, waveform acceleration time histories recorded at the surface elements were examined for evidence of boundary-related reflections. Waveforms obtained from ILM and IDM at each inclination angle are shown in Figure 5, where each surface waveform has been normalized by its maximum amplitude. It can be observed that the ILM produced relatively minimal boundary reflections across all tested inclination angles. Its staggered input application and fixed horizontal domain allowed the inclined wavefront to propagate smoothly, with only minor spurious energy reflections initiating from the free-field boundaries. In contrast, the IDM generated substantially larger reflections than the ILM when the inclination angle exceeded 5°. These reflections occurred when the vertically-applied base motion reached the inclined/rotated free surface, and scattered back into the domain along the free-field boundaries, producing pronounced interference patterns at higher inclinations (15º, 30°). Such issues would only be exacerbated in 3D, where multidirectional scattering and the larger free surface further amplify boundary-related artifacts.

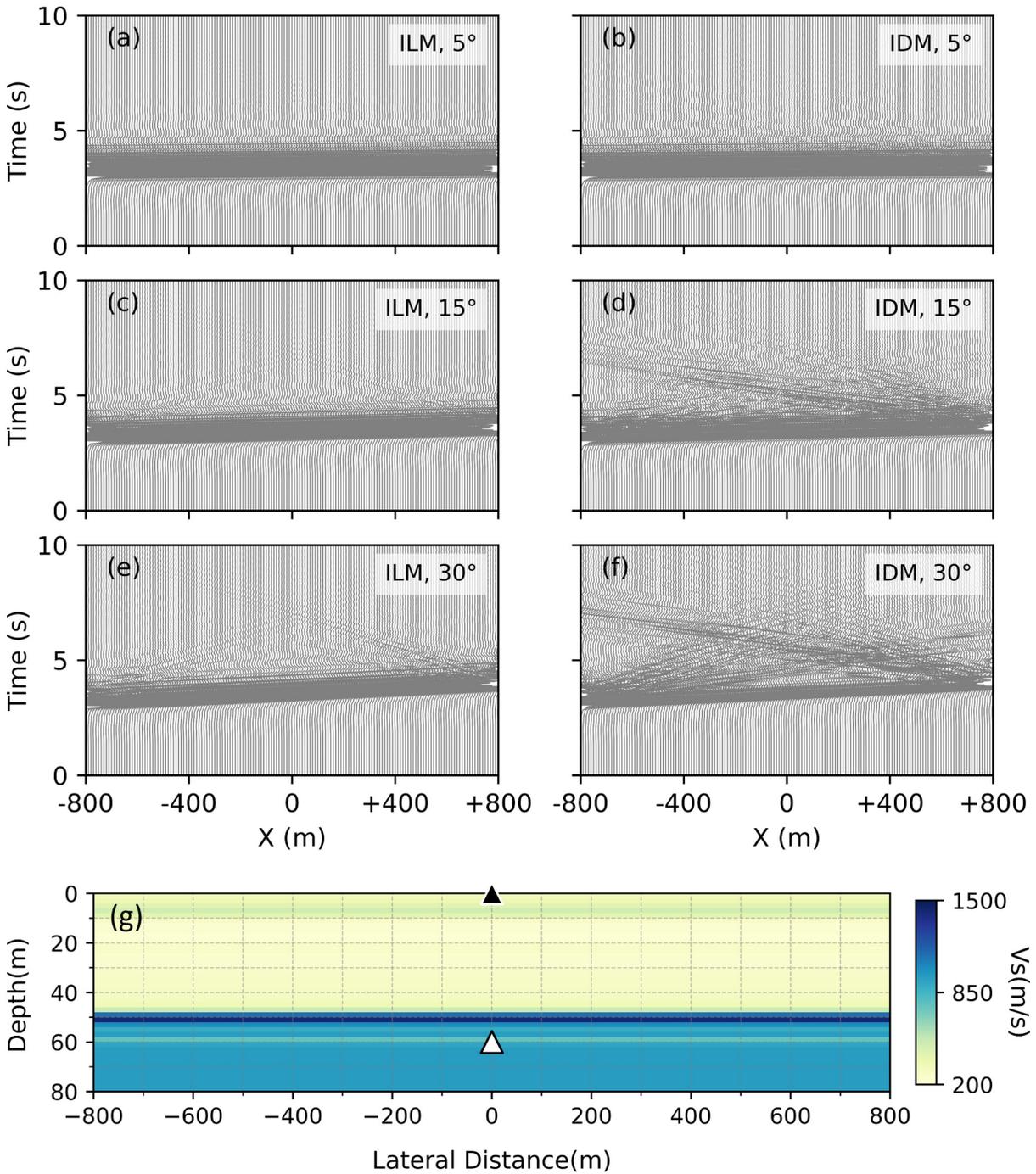

**Figure 5.** Comparison of inclined-wave simulations using the ILM and the IDM for inclination angles of 5°, 15°, and 30°. Shown are: (a), (c), and (e) simulated surface accelerations obtained using the ILM; (b), (d), and (f) simulated surface accelerations obtained using the IDM; and (g) the laterally-homogeneous DPDA 2D cross-section used in these tests. Solid and hollow triangles represent surface and downhole sensors at DPDA, respectively.

Although the FLAC3D documentation (Itasca Consulting Group, 2023) recommends the IDM for applying inclined waves, this recommendation is most applicable to relatively small models, where the required domain expansion remains manageable, and to nonlinear simulations in which gravitational effects can be readily accommodated by rotating the gravity field. For the kilometer-scale pseudo-3D subsurface models used in this study, however, implementation of the IDM requires a substantially enlarged computational domain and leads to a significant increase in computational cost. In addition, the IDM clearly results in stronger boundary reflections under these conditions, making it less suitable for large-scale site response analyses of the type considered here. Given that the ILM consistently produced cleaner and more stable results while avoiding the severe computational demands associated with the IDM, it was adopted as the preferred approach for applying inclined waves. Accordingly, all subsequent 2D and 3D GRAs in this study were performed using the ILM.

### 4.4  ILM for 3D GRA

For 3D GRAs, the ILM accommodates both the inclination and azimuth of the incident wave, allowing the propagation direction to vary in both the vertical and horizontal planes. To represent this behavior, a spatially varying time delay is applied across the entire 2D base of the 3D model, ensuring that the incident wavefront enters the domain with the prescribed inclination angle $\theta$ and azimuthal angle $\alpha$. As illustrated in Figure 6, an inclined wave propagating in the direction $\hat{w}(\theta, \alpha)$, i.e., at an azimuth $\alpha$ measured clockwise from north and an inclination angle $\theta$ measured clockwise from vertical, intersects the horizontal model base along a line defined by the unit vector $\hat{h}(\alpha)$. For the center of the $n^{th}$ base element, located at point $P_n(x_n, y_n, 0)$, the effective arrival time of the wavefront is governed by the projection of $P_n$ onto $\hat{h}(\alpha)$, denoted as $proj_{\hat{h}(\alpha)}(P_n)$ in Figure 5a. This projection defines the input offset, which is used to compute the lag distance and subsequently converted to a time delay using the halfspace $V_s$, as given in Equation (8). The delayed application of the input motion at each base element causes the wavefront to sweep across the model base with the correct inclination and azimuth, thereby producing a physically consistent 3D inclined plane wave.

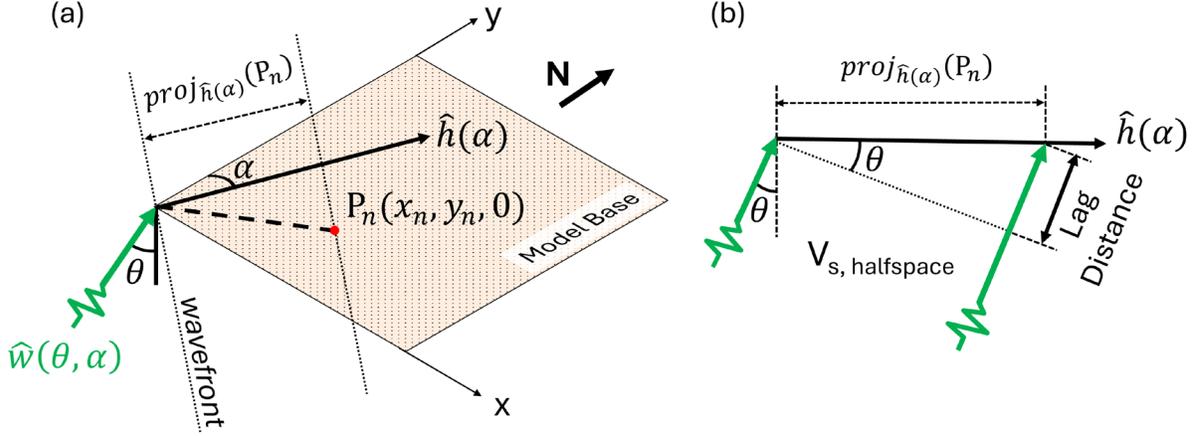

**Figure 6.** Schematic illustration of the 3D ILM. Shown are: (a) geometry of an inclined plane wavefront with inclination $\theta$ and azimuth $\alpha$, and the projection of a model-base point $P_n(x_n, y_n, 0)$ onto the wavefront-propagation unit vector $\hat{h}(\alpha)$; and (b) side view illustrating the corresponding lag distance used to compute the time delay applied at each base element.

An inclined wave propagating in the direction $\hat{w}(\theta, \alpha)$ intersects the plane of the horizontal model base along a line defined by the unit vector $\hat{h}(\alpha)$:

$$\hat{h}(\alpha) = (\sin\alpha, \ \cos\alpha, \ 0) \quad (6)$$

The projection of point, $P_n(x_n, y_n, 0)$, onto the unit vector $\hat{h}(\alpha)$ is given by:

$$proj_{\hat{h}(\alpha)}(P_n) = x_n \sin\alpha + y_n \cos\alpha \quad (7)$$

The input lag, $\Delta t_n$, can be defined by:

$$\Delta t_n = \frac{\text{Lag Distance}}{V_{s,\text{halfspace}}} = \frac{proj_{\hat{h}(\alpha)}(P_n) \times \sin\theta}{V_{s,\text{halfspace}}} \quad (8)$$

From Equations (7) and (8), we can write:

$$\Delta t_n = [x_n \sin\alpha + y_n \cos\alpha] \frac{\sin\theta}{V_{s,\text{halfspace}}} \quad (9)$$

Equation (9) represents the input lag, $\Delta t_n$, that can be applied to the input motion in order to simulate the staggered arrival of inclined waves at any azimuth at the base of the 3D model. In addition to staggering the input motion, the incident inclined wave field must be decomposed into two horizontal and one vertical components. However, the formulation of inclined synthetic shear waves in 3D is inherently more complex than in 2D. In the 2D ILM formulation, the inclined shear wave is confined to the vertical plane of the cross-section, and no out-of-plane motion is introduced, resulting in a relatively straightforward decomposition and application of the input motion. In 3D, however, the incident shear wave consists of two polarized components: a shear wave polarized within an inclined horizontal plane ($S_{H,inc}$) and a shear wave polarized at an angle within a vertical plane ($S_{V,inc}$). Consequently, the scaling factors required to distribute the input motion across the model base must account for both shear-wave components, making the formulation more involved, as the prescribed inclination and azimuth jointly control their relative contributions. This added complexity underscores the importance of a carefully defined 3D ILM formulation to ensure that the intended inclined shear-wave field is introduced without inadvertently generating undesired wave modes or directional artifacts. Figure 7 illustrates the base of the 2D model extracted from the 3D domain, excited by a shear wave propagating along the unit vector $\hat{w}(\alpha, \theta)$. This inclined shear wave needs to be resolved in two mutually orthogonal directions: a horizontally polarized shear component confined to an inclined horizontal plane ($\hat{S}_{H,inc1}$ and $\hat{S}_{H,inc2}$) and a vertically polarized shear component inclined within a vertical plane ($\hat{S}_{V,inc1}$ and $\hat{S}_{V,inc2}$). The subscripts 1 and 2 indicate positive and negative particle-motion directions for each polarization, respectively.

The unit vector, $\hat{w}(\alpha, \theta)$, can be defined as:

$$\hat{w}(\alpha, \theta) = \hat{h}(\alpha) \sin\theta + \hat{z} \cos\theta \tag{10}$$

Using Equation 6 for $\hat{h}(\alpha)$ and substituting $(0, 0, 1)$ for $\hat{z}$, we get:

$$\hat{w}(\alpha, \theta) = (\sin\theta \sin\alpha, \sin\theta \cos\alpha, \cos\theta) \tag{11}$$

Equation (11) defines the unit vector describing the propagation direction of the inclined wave in 3D. As shown in Figure 7, at the instant the inclined wave reaches the model base, $\hat{S}_{H,inc1}$ is orthogonal to both the propagation-direction unit vector $\hat{h}(\alpha)$ and the vertical unit vector $\hat{z}$. Accordingly, using the right-hand rule for vector cross products, the unit vector associated with $\hat{S}_{H,inc1}$ is given by,

$$\hat{S}_{H,inc1} = \frac{\hat{h}(\alpha) \times \hat{z}}{|\hat{h}(\alpha) \times \hat{z}|} = (\cos\alpha, -\sin\alpha, 0) \tag{12a}$$

Similarly, for the negative polarization,

$$\hat{S}_{H,inc2} = \frac{\hat{z} \times \hat{h}(\alpha)}{|\hat{z} \times \hat{h}(\alpha)|} = (-\cos\alpha, \sin\alpha, 0) \tag{12b}$$

From Figure 7, it can also be seen that the unit vector $\hat{S}_{V,inc1}$ is perpendicular to the propagation direction of the inclined wave, $\hat{w}(\alpha, \theta)$. In addition, $\hat{S}_{V,inc1}$ is essentially orthogonal to $\hat{S}_{H,inc1}$. Accordingly, the unit vector associated with $\hat{S}_{V,inc1}$ can be expressed as:

$$\hat{S}_{V,inc1} = \frac{\hat{w}(\alpha, \theta) \times \hat{S}_{H,inc1}}{|\hat{w}(\alpha, \theta) \times \hat{S}_{H,inc1}|} = (\cos\theta \sin\alpha, \cos\theta \cos\alpha, -\sin\theta) \tag{13a}$$

Likewise, for the negative polarization,

$$\hat{S}_{V,inc2} = \frac{\hat{S}_{H,inc2} \times \hat{w}(\alpha, \theta)}{|\hat{S}_{H,inc2} \times \hat{w}(\alpha, \theta)|} = (-\cos\theta \sin\alpha, -\cos\theta \cos\alpha, \sin\theta) \tag{13b}$$

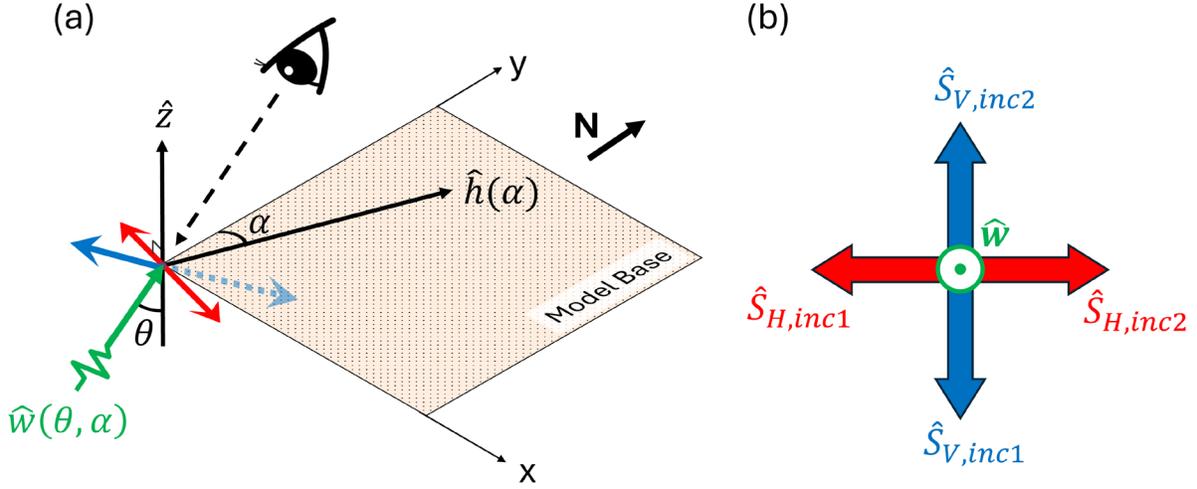

**Figure 7.** Schematic illustration of inclined shear-wave incidence and polarization. Shown are: (a) Incident-wave geometry illustrating the inclination angle $\theta$, azimuth $\alpha$, and the direction of propagation defined by the unit vector $\hat{h}(\alpha)$. The eye symbol indicates the observer's viewing direction used to define the polarization sense of the incident wave. (b) Decomposition of the inclined shear wave, as viewed from the observer's perspective, into two orthogonal polarized components: a horizontally polarized shear wave confined to an inclined horizontal plane ($\hat{S}_{H,inc1}$ and $\hat{S}_{H,inc2}$) and a vertically polarized shear component inclined within a vertical plane ($\hat{S}_{V,inc1}$ and $\hat{S}_{V,inc2}$). Subscripts 1 and 2 denote positive and negative particle-motion directions, respectively. The green dot at the center indicates that, at the instant shown, the inclined wave is propagating toward the observer.

These unit vectors define the polarization of the incident shear wave, and their components along the $\hat{i}$, $\hat{j}$, and $\hat{k}$ components specify the scaling factors associated with x, y, and z coordinate directions. Multiplication of the polarization unit vector by the prescribed input motion amplitude yields the actual motion components applied in each coordinate direction. The total scaling factors, $f_x$, $f_y$, and $f_z$, in each direction are obtained by summing the contributions from the positively polarized components, $\hat{S}_{H,inc1}$ and $\hat{S}_{V,inc1}$, as defined in Equation (14). The negatively polarized components, $\hat{S}_{H,inc2}$ and $\hat{S}_{V,inc2}$, represent particle motion of equal magnitude but opposite sign and therefore do not need to be treated separately, as their effects are inherently accounted for through the changing polarity of the applied input motion.

$$f_x = \cos\alpha + \cos\theta\ \sin\alpha \tag{14a}$$

$$f_y = -\sin\alpha + \cos\theta\ \cos\alpha \tag{14b}$$

$$f_z = -\sin\theta \tag{14c}$$

Using Equations (2) and (14), the equivalent shear-stress input for a 3D inclined wave can be expressed as,

$$\tau_x = \tau \cdot f_x = -2\rho V_s P_s (\cos\alpha + \cos\theta\ \sin\alpha) \tag{15a}$$

$$\tau_y = \tau \cdot f_y = -2\rho V_s P_s (\sin\alpha - \cos\theta\ \cos\alpha) \tag{15b}$$

$$\tau_z = \tau \cdot f_z = 2\rho V_s P_s \sin\theta \tag{15c}$$

Finally, the equivalent shear-stress input defined in Equation (15) are applied simultaneously at a given element along the model base with an element-specific time lag, $\Delta t_n$, as defined in Equation (9). This staggered application generates an incident wave governed by both the inclination and azimuthal angles, causing the wavefront to propagate across the model base in a physically realistic manner.

### 4.5  ILM for 2D/3D GRAs using real ground motion

While this study employs synthetic waves for parametric analysis, it is common in GRAs to use recorded ground motions. Ground motions recorded at permanently installed seismic stations are measured in three orthogonal components (two horizontal and one vertical). These components reflect a superposition of multiple wave types and propagation directions and therefore cannot be uniquely associated with individual horizontally or vertically polarized shear waves ($S_H$ and $S_V$, respectively) or compressional ($P$) waves, even for vertically incident motion. This complexity is further enhanced for inclined incidence due to $P - S_V$ mode conversions, radiation effects, and local heterogeneity. As a result, the treatment of recorded ground motions for inclined-wave simulations requires explicit modeling assumptions.

Unlike synthetic inclined waves, which are prescribed by construction to propagate in a specified direction and therefore require decomposition into x, y, and z components, recorded ground motions are already measured in orthogonal directions. As such, they inherently represent projections of incident wavefields arriving from unknown azimuthal and inclination angles. Consequently, additional decomposition is neither necessary nor physically meaningful. One possible modeling approach is to interpret the recorded vertical component as representing vertical particle motion associated with an inclined shear wave and to include it as part of the shear-wave input. However, recorded vertical motions are typically dominated by $P$-wave energy, and these contributions cannot be uniquely separated from the inclined shear-wave response. Therefore, when the objective is to isolate the effects of inclined shear-wave incidence, it is recommended to apply only the horizontal components of motion at the model base.

For 3D GRAs, the two horizontal components can be applied directly along their respective directions. In contrast, for 2D GRAs, the two horizontal components should be rotated into the orientation of the cross-section and combined to generate a single horizontal shear-wave input. In both cases, the ground motion must be applied as a spatially staggered input across the model base to reproduce the target incident direction. For 2D GRAs, the required time lag, $\Delta t_n$, depends on the inclination angle, $\theta$, as defined in Equation (3). For 3D GRAs, the time lag depends on both the inclination angle, $\theta$, and the azimuthal angle, $\alpha$, as defined in Equation (9).

### 5.  Results and Discussions

To evaluate the effects of inclined and azimuthally varying incident waves on site response at DPDA, input Ricker wavelets were propagated through the 2D and 3D numerical models using FLAC3D. Accelerations were recorded at the surface and at the depth of the deepest downhole sensor, and TTFs were computed in the same manner as the ETFs described in Section 2. To ensure accurate comparisons with ETFs, all FAS used in this study were smoothed using the Konno and Ohmachi (1998) method with a bandwidth coefficient (b-value) of 75, which is a more moderate smoothing compared to the b-value of 40 which is typically used for H/V processing (e.g., Molnar et al., 2022; Cox et al., 2020). It is important to note that both the ETFs and TTFs are considered as "within" transfer functions, which include the combined effects of upgoing and downgoing waves at the deepest sensor. To maintain this consistency, TTFs were computed using the simulated motion at the sensor depth rather than the input motion applied at the model base, which contains only upgoing energy. All numerical models were therefore extended below the deepest downhole sensor so that the input motion could be applied without influencing the sensor-level response.

## 5.1   Effects of inclined waves on 2D GRA

The effects of inclined incident waves on 2D site response at DPDA were evaluated using the most heterogeneous cross-section extracted from the pseudo-3D $V_s$ model along Az = 165° (cross-section A–A′; see Figures 1b and 1c). A range of inclination angles was modeled to assess both the sensitivity and directional dependence of TTFs, where positive angles ($+\theta$) represent waves arriving from the A direction (approximately north) and negative angles ($-\theta$) represent waves arriving from the A′ direction (approximately south). In the context of 2D GRAs, this $+\theta$ versus $-\theta$ comparison is effectively equivalent to examining two opposite azimuthal directions within the same vertical plane, because changing the sign of $\theta$ reverses the horizontal propagation direction along the cross-section. This provides a convenient way to explore azimuthal sensitivity within a 2D framework, although for clarity we continue to refer to them simply as positive and negative inclination angles. The full suite of simulations included $\theta = 0°$, $\pm 5°$, $\pm 10°$, $\pm 15°$, $\pm 30°$, $\pm 45°$, and $\pm 60°$, enabling comparison across both small and large deviations from vertical incidence.

The resulting TTFs for the A-direction ($+\theta$) and A′-direction ($-\theta$) waves are shown in Figures 8a and 8b, respectively. Because the 2D GRAs were carried out using a 2-m cell size, corresponding to $f_{s,2m} = 11.2$ Hz (indicated by the green dash–dot line in Figure 8), all three higher-mode features of the ETFs fall within the resolved frequency range and are therefore suitable for comparison. For small inclinations ($|\theta| \leq 15°$), the TTFs retain shapes quite similar to the vertical-incidence case. The fundamental peak frequency remains largely unchanged, exhibiting only minor shifts to slightly higher frequencies on the order of a few tenths of a hertz. Amplitude changes in $f_0$ are also modest, with only a slight reduction for small positive angles ($+5°$ to $+15°$) and virtually no amplitude change for the corresponding negative angles ($-5°$ to $-15°$). The primary differences in the TTFs at small inclination angles are the appearance of secondary peaks and troughs that occur in the vicinity of the ETF trough between $f_0$ and $f_1$, which result from wave interference patterns. However, these effects are minor and overall, the results indicate that small incident angles do not meaningfully reduce or broaden the TTF $f_0$ peak or disrupt the constructive interference patterns responsible for amplitude overpredictions at DPDA. As inclination increases, the TTFs become increasingly sensitive to how the incoming wavefront interacts with the dipping impedance contrast. For larger inclinations ($|\theta| \geq 30°$), the TTFs experience clear, systematic changes. The $f_0$ peak consistently shifts toward higher frequencies, and its amplitude decreases as inclination increases, except for the $\theta = -30°$ case, in which a slight amplitude increase is observed. In contrast, the $f_1$ peak shifts slightly toward lower frequencies while its amplitude increases. These trends reflect stronger scattering and altered propagation paths as the inclined wavefront intersects the till–clay interface at more oblique angles. The effects are more pronounced for the A′-direction ($-\theta$) waves, as the waves approach the dipping interface more nearly perpendicular and, therefore, experience stronger impedance-driven scattering. A-direction ($+\theta$) waves, which propagate more parallel to the dip, exhibit relatively smaller peak-frequency shifts and more gradual amplitude changes. Comparison of Figures 8a and 8b shows that, although both directions display similar qualitative patterns, the magnitude of these inclination effects depends strongly on whether waves arrive from the A or A′ direction.

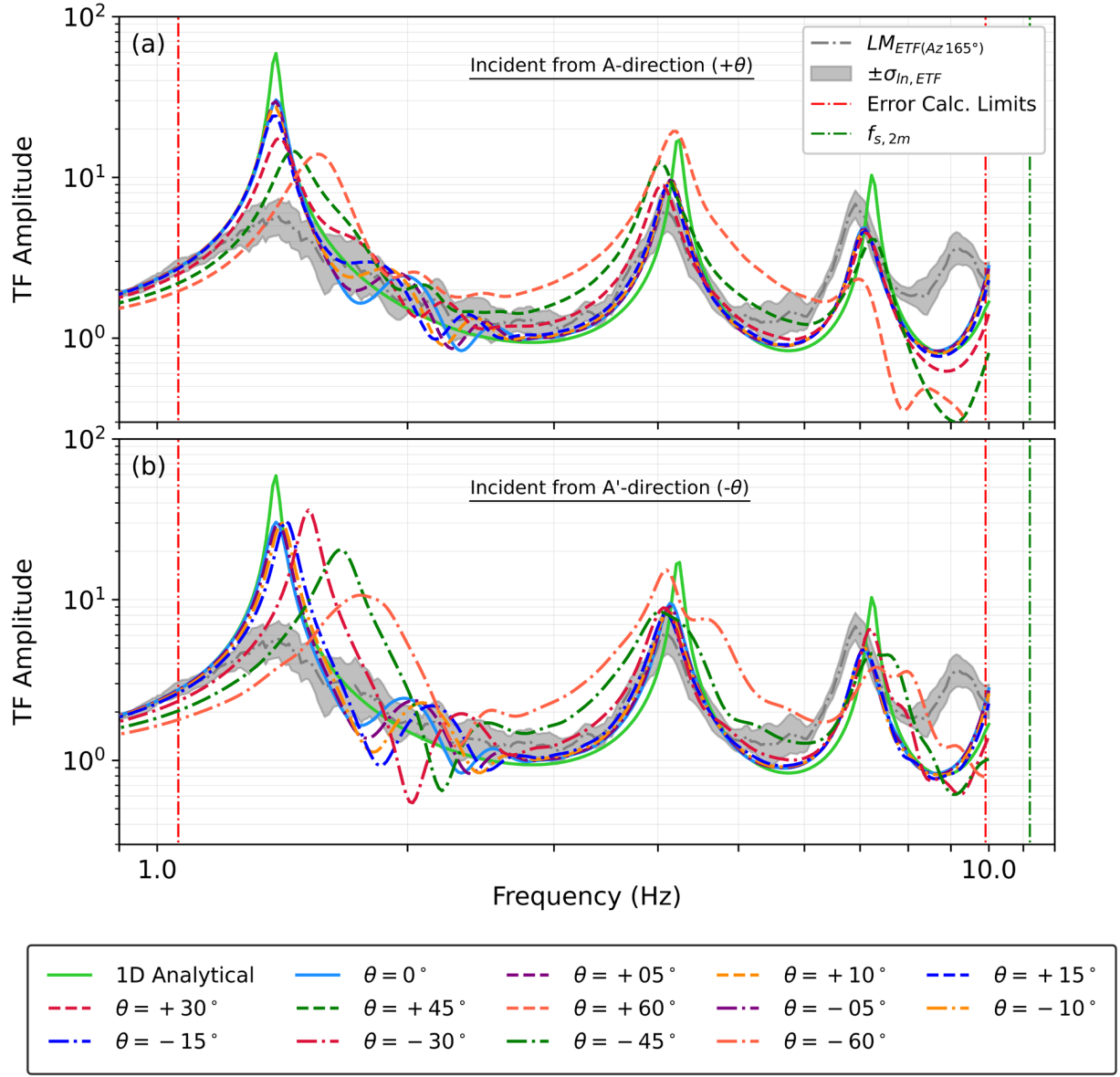

**Figure 8.** Site response prediction results from 2D GRAs conducted along the Az165° DPDA cross-section using input motions with varying inclination angles. Shown are: (a) comparisons between TTFs and ETFs for waves incident from the A-direction (+$\theta$), and (b) the same comparisons for waves incident from the A′-direction (−$\theta$). The A- and A′-directions (refer to Figure 1b) correspond approximately to north and south, respectively.

Figure 9 presents two quantitative goodness-of-fit metrics used to compare TTFs and ETFs: the Pearson correlation coefficient ($r$) and the transfer function misfit ($m_{TF}$). The correlation coefficient quantifies how closely the TTFs align with the median ETF, with higher values indicating better predictions. A perfect match yields $r = 1.0$, and values above 0.6 are considered good (Thompson et al., 2009). The $m_{TF}$ measures the average deviation of the TTFs from the median ETF in units of standard deviation. Although there are no definitive thresholds for acceptable $m_{TF}$ value, due to its dependence on $\sigma_{\ln ETF}$ which varies with frequency and across sites, lower $m_{TF}$ values are preferred. A value of $m_{TF} = 1.0$ indicates that the TTF lies within one standard deviation of the median ETF over the frequency range considered. For more information on these metrics and their application in assessment of site response predictions, readers are referred to Teague et al. (2018). For the 2D GRAs, $r$ and $m_{TF}$ values were calculated over the frequency range extending from the half-amplitude point preceding the ETF $f_0$ peak to the half-amplitude point following the $f_3$ peak, or up to 10 Hz, whichever was smaller, following the procedures outlined by Hallal and Cox (2021b). These error-metric calculation limits are indicated by red dash–dot lines in Figure 8. The 2D cross-section is shown in Figure 9a to help visualize wave-propagation directions relative to the dipping bedrock structure. Figure 9b shows the $r$ and $m_{TF}$ values for larger inclination intervals across the range of $\theta = 0°$ to $\pm 60°$, while Figure 9c shows the corresponding values for smaller inclination intervals across the range of $\theta = 0°$ to $\pm 15°$. For the A-direction waves, $r$ values increase slightly with increasing wave inclinations up to approximately $+30°(r \approx 0.75)$, indicating slightly better agreement between the TTFs and ETF. However, the $m_{TF}$ values increase slightly over this same range, indicating slightly poorer agreement. The $r$ values decrease sharply between $+30°$ and $+60°$, and the $m_{TF}$ values increase significantly over this same range, both metrics indicating significantly poorer agreement between the TTFs and ETF. For inclined waves propagating from the A′-direction (i.e., negative inclinations), $r$ values decrease for even modest increases in inclination angle and drop sharply for larger inclinations. The $m_{TF}$ values increase in a similar pattern for these negative inclination angles, indicating poorer agreement between the TTFs and ETF.

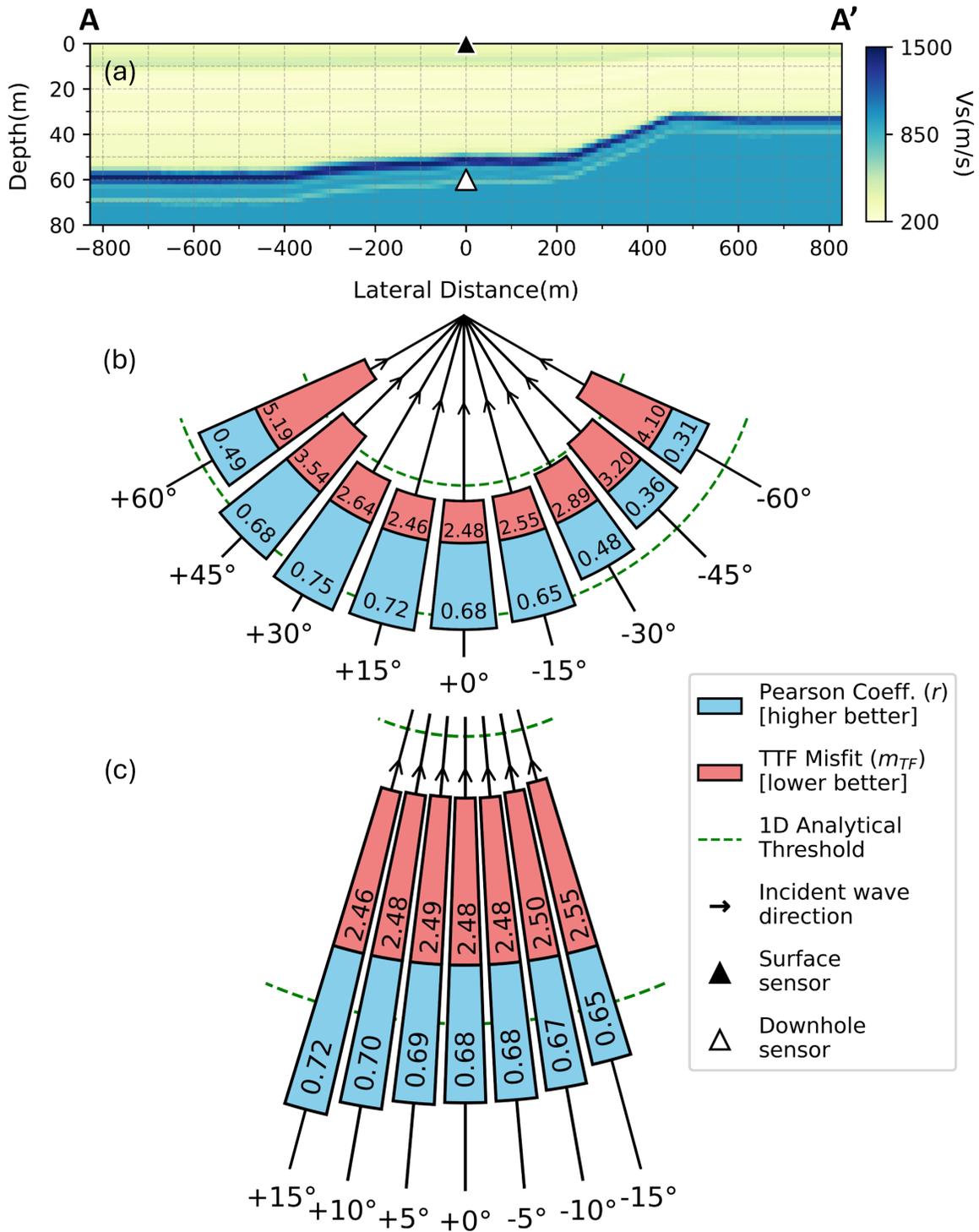

**Figure 9.** Quantitative comparison of site response prediction results from 2D GRAs conducted along the Az165° DPDA cross-section using input motions with varying inclination angles. Shown are: (a) the Az165° cross-section used in the analyses (see A–A' in Figure 1b); (b) goodness-of-fit metrics, $r$ and $m_{TF}$, for larger inclination angles (0°, ±15°, ±30°, ±45°, and ±60°); and (c) the corresponding statistics for smaller inclination angles (0°, ±5°, ±10°, and ±15°).

In summary, the results in Figure 8 and Figure 9 qualitatively and quantitatively demonstrate that both the magnitude and direction of wave inclination exert a meaningful influence on the 2D site response at DPDA when the waves are inclined at angles greater than $+/- 30°$. However, TTFs calculated from waves with these larger incidence angles do not improve agreement between the TTFs and ETF, indicating it is unlikely that waves with these larger incidence angles are present in the small-strain ground motions recorded at DPDA. Furthermore, the shift in $f_0$ observed in TTFs at larger inclination angles (refer to Figure 8) can be attributed to the fact that inclined waves sample a different effective soil column than vertically incident waves, resulting in a change in the apparent resonance frequency. The absence of comparable peak-frequency shifts in the observed ETFs indicates that the ground motions used in this study did not contain strongly inclined incident waves. Inclined waves with modest positive inclination angles (up to $+15°$) slightly improve agreement between TTFs and the ETF in terms of both $r$ and $m_{TF}$, but do not significantly reduce the over-amplification predicted by the TTFs at $f_0$ (refer to Figure 8). In aggregate, these observations suggest that the relatively low $f_0$ amplification observed in the ETFs may be somewhat influenced by moderately inclined incident waves arriving from the north–northwest direction, which corresponds to a notable concentration of epicentral azimuths among the ground motions used to compute the ETFs in this study (refer to Figure 2). Nevertheless, inclined waves modeled in 2D do not appear to be responsible for the significant amplitude mismatch between the TTFs and ETFs at $f_0$. Notably, the site response sensitivity to modeling wave inclination at DPDA is stronger than the behavior reported by Eskandarighadi et al. (2026) at the TIDA downhole array, where inclination effects were less pronounced than those shown for DPDA. Still, this parametric study indicates that wave inclination effects alone cannot explain the significant differences between the TTF and ETF amplitudes at the fundamental mode.

## 5.2   *Effects of inclined waves on 3D GRA*

The influence of inclined incident waves on 3D site response at DPDA was examined using the full pseudo-3D $V_s$ model, with simulations performed at inclination angles of 0°, 5°, 10°, 15°, 30°, and 60° for waves impinging from an azimuth of 165°, consistent with the A-direction used for positive incidence angles in the 2D GRAs conducted on the Az165 cross-section (refer to Figure 1b). As discussed in Section 2, conducting numerous 3D GRAs with the original 25.6-million-element model discretized using 2-m cubic elements was computationally impractical; therefore, the pseudo-3D $V_s$ model was re-discretized using 5-m cubic elements, reducing the model size to approximately 1.64 million elements. This coarser discretization reduces the maximum resolvable frequency of the 3D simulations to $f_{s,5m} = 4.48$ Hz, and the interpretation of 3D GRA results is therefore limited to frequencies below this threshold. Although the 5-m discretization satisfies the adopted resolution criterion of at least 10 elements per wavelength only up to $f_{s,5m}$, results are nevertheless presented up to 10 Hz to enable qualitative comparison with the observed ETFs over a broader frequency range. At 10 Hz, the 5-m mesh still provides approximately 4–5 elements per shortest propagating wavelength, which, while insufficient for the best quality results, retains qualitative information on the overall shape and trends of the TTFs. Accordingly, results above $f_{s,5m}$ are shown for contextual comparison only and should be interpreted with appropriate caution.

Figure 10a presents the TTFs obtained from 3D GRAs performed using a range of inclination angles. Because the 3D GRAs produce two orthogonal horizontal acceleration components, the simulated EW- and NS-acceleration time histories at both the downhole and surface sensors were combined and rotated into the Az = 165° direction, and the TTFs were computed using these directionally-resolved acceleration records to enable direct comparison with the Az165 ETFs shown in Figure 2e. For smaller inclination angles ($\theta \leq 15°$), the frequency and amplitude of the $f_0$, $f_1$, and $f_2$ peaks remain largely unchanged. At $\theta = 30°$, a slight positive shift in the $f_0$ frequency is observed, accompanied by a modest negative shift in the $f_1$ frequency, while the amplitude of the $f_0$ peak is noticeably reduced. For $\theta = 60°$, the shift in the $f_0$ frequency becomes much more pronounced, with an accompanying reduction in the $f_0$-peak amplitude. In contrast, the amplitude of the $f_1$-peak increases significantly. These results are similar to the trends observed for larger positive inclination angles in the 2D GRAs (refer to Figure 8). More in-depth comparisons between the 2D and 3D GRA results are provided below in Section 5.3.

Figure 10b presents the quantitative TTF-to-ETF goodness-of-fit metrics, $r$ and $m_{TF}$, computed over the frequency range extending from the half-amplitude point preceding the ETF $f_0$ peak to the half-amplitude point following the $f_1$ peak, as indicated by the red dash–dot lines in Figure 10a. The maximum resolvable frequency of the 3D model, $f_{s,5m}$, is also shown using a blue dashed line. In Figure 10b, the $r$ and $m_{TF}$ values are plotted against one another. High $r$ values and low $m_{TF}$ values indicate good fits in the upper left of the figure, while low $r$ values and high $m_{TF}$ values indicate poor fits in the lower right of the figure. The qualitative trends observed in Figure 10a are well reflected in the corresponding $r$ and $m_{TF}$ values. Smaller inclination angles ($\theta \leq 15°$) yield very similar performance, with $r$ values ranging from 0.78 to 0.79 and $m_{TF}$ values between 1.6 and 1.7, which represent a substantial improvement relative to the 1D analytical TTF ($r = 0.66$, $m_{TF} = 2.35$). The $\theta = 30°$ case provides the best overall fit, with $r = 0.82$ and $m_{TF} = 1.52$, whereas the $\theta = 60°$ case performs the poorest ($r = 0.54$, $m_{TF} = 3.23$), even underperforming the 1D TTF, primarily due to the large shift in the $f_0$.

While modeling wave inclination in 3D resulted in slight improvements in TTF fits (for inclination angles up to approximately 30°), it did not produce agreement with the lower $f_0$ amplitudes observed in the ETFs. The 3D simulations with inclined wavefields therefore serve primarily as a parametric test of whether wave inclination could explain the lower ETF $f_0$ amplitudes, and they show that this is unlikely. Only relatively large inclination angles produce a noticeable reduction in TTF amplitudes; however, similar to the 2D GRA results, these cases are accompanied by shifts in the fundamental frequency, reflecting sampling of a different effective soil column. This coupled behavior is not consistent with ETF observations at DPDA. These findings are broadly consistent with those reported by Dawadi et al. (2026a) for the I15DA site, who similarly observed that inclined wavefields do not resolve the amplitude discrepancy for small to moderate inclination angles and lead to poorer agreement at steeper angles due to peak shifts. Although the response at DPDA appears more sensitive to inclination, likely due to its greater lateral heterogeneity and more complex dipping stratigraphy, the overall results indicate that inclination effects alone cannot account for the mismatch between TTFs and ETFs.

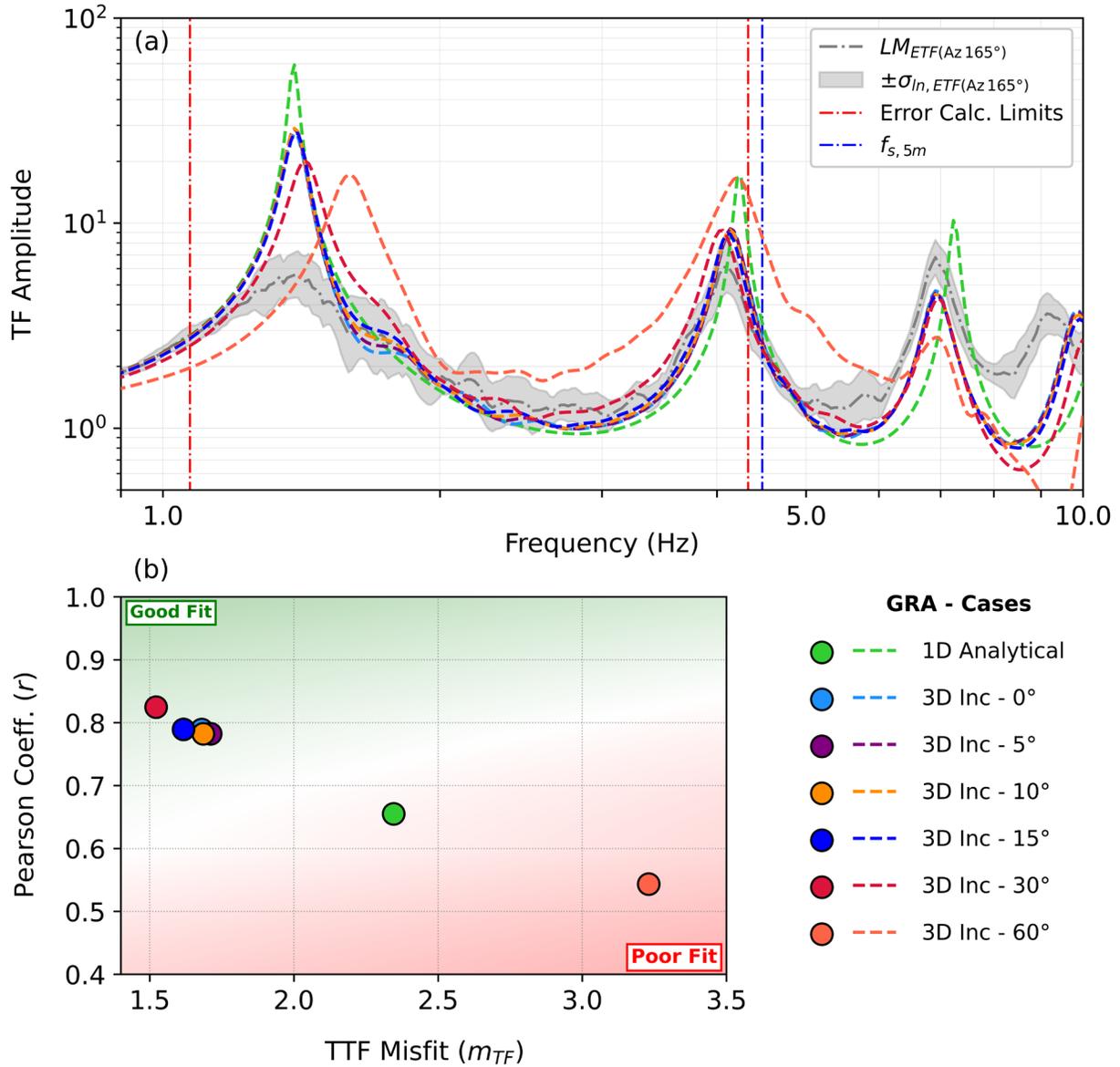

**Figure 10.** Site response prediction results from 3D GRAs at DPDA conducted using input motions along an azimuthal direction of Az165° and varying inclination angles. Shown are: (a) comparisons between TTFs and ETFs; and (b) goodness-of-fit metrics, $r$ and $m_{TF}$, for each comparison.

### 5.3  Comparison of 2D and 3D GRA

Figure 11 presents a comparison of the TTFs obtained from 2D and 3D GRAs at DPDA conducted using input motions along Az165° (A-direction for the 2D analyses) and varying inclination angles. Because the 3D GRAs were performed using the re-discretized 5-m pseudo-3D model, the frequency range over which 2D and 3D TTFs are compared with the observed ETFs is controlled by the resolution of the 3D simulations, which represent the most restrictive case. Accordingly, quantitative TTF–ETF comparisons and goodness-of-fit metrics are evaluated only up to $f_{s,5m}$, ensuring a consistent and numerically reliable basis for comparing 2D and 3D results. From Figure 11, it is evident that, for vertical incidence, both 2D and 3D GRAs provide a clear improvement over the 1D analytical solution. For the $\theta = 0°$ case, the 2D GRA increases the Pearson correlation from $r = 0.66$ for the 1D solution to $r = 0.77$, while the 3D GRA yields a slightly higher value of $r = 0.79$. Both approaches are also associated with reduced misfit values relative to the 1D case, indicating that accounting for lateral heterogeneity significantly improves agreement with the observed ETFs. Comparing the 2D and 3D results for vertical incidence ($\theta = 0°$), the additional improvement gained by moving from 2D to 3D is modest. Although the 3D GRA produces a slightly higher correlation ($r = 0.79$ versus $r = 0.77$) and lower misfit ($m_{TF} = 1.68$ versus 1.88), the overall $f_0$ peak characteristics remain similar for both cases. Some differences are observed in the spectral shape of the troughs: the 2D GRA TTFs consistently exhibit secondary peaks near ~2 Hz, whereas these features are absent in the 3D GRA results. This behavior reflects the fact that, in 3D GRAs, scattering is not confined to a single vertical plane, allowing wave energy to be distributed more broadly throughout the three-dimensional domain. This trend is observed not only for the vertical-incidence case but across all modeled inclination angles. Introducing inclined incidence affects both 2D and 3D GRAs in a broadly similar manner. For small to moderate inclination angles ($\theta = 5° - 15°$), both modeling approaches show improved agreement with the ETFs, with optimum performance occurring at $\theta = 15°$. At this inclination, the 2D GRA yields the highest correlation ($r = 0.82$), while the 3D GRA produces the lowest misfit ($m_{TF} = 1.62$), indicating comparable sensitivity to inclination in both dimensions. At steep inclination ($\theta = 60°$), agreement deteriorates markedly for both approaches, with reduced correlation and substantially increased misfit caused by large shifts in $f_0$. This degradation is more pronounced in the 3D GRA results.

Overall, these results indicate that 2D and 3D GRAs exhibit similar response to wave inclination at DPDA, with modest inclination angles slightly reducing $f_0$-peak amplitude and steep incidence leading to $f_0$ shifts. They further suggest that the added computational complexity of full 3D modeling does not provide a substantial improvement over 2D GRAs at DPDA, and that much of the benefit relative to 1D behavior is already captured within the 2D framework when the azimuthal cross-section with the greatest spatial variability is used.

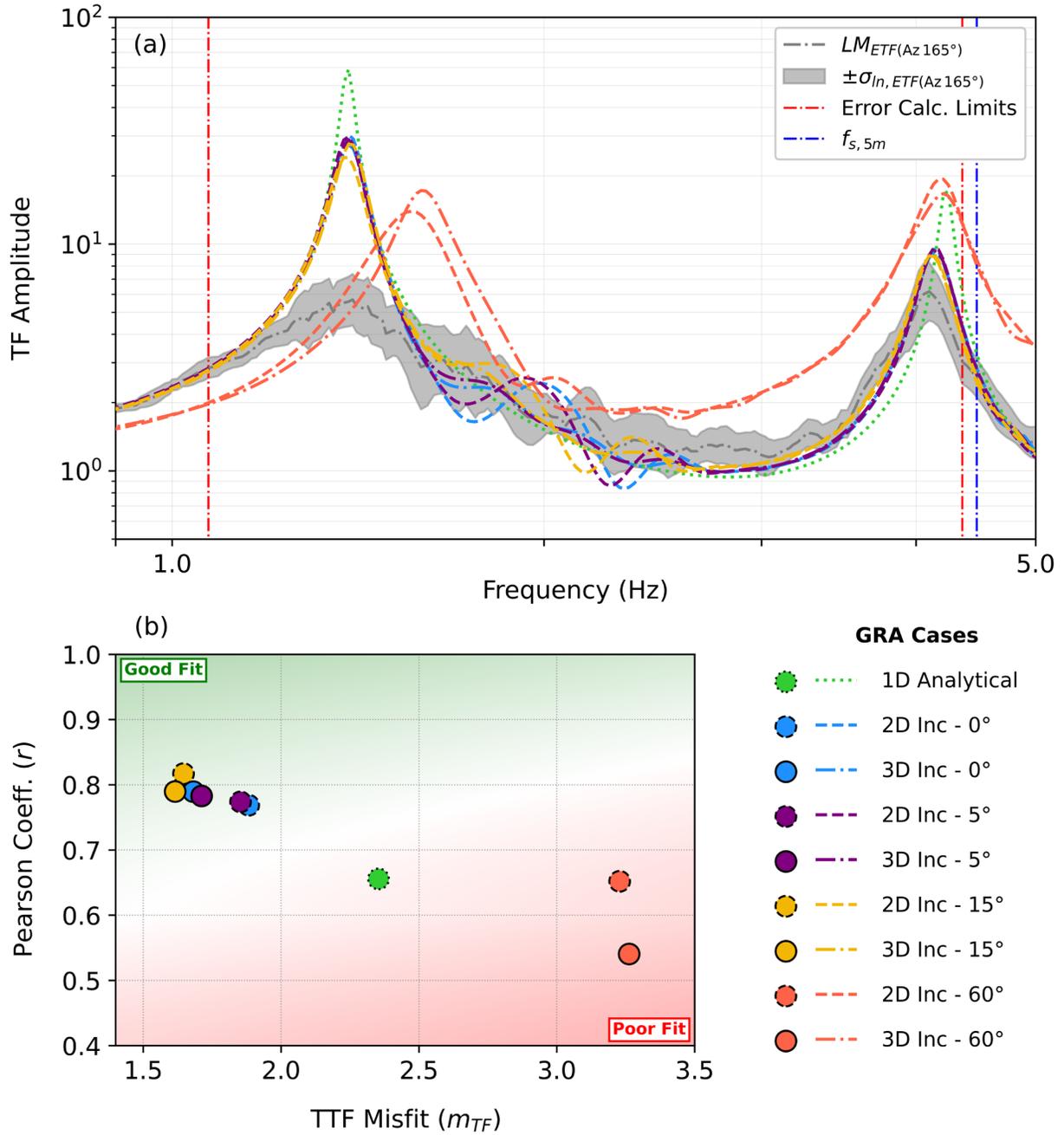

**Figure 11:** Site response prediction results from 2D and 3D GRAs at DPDA conducted using input motions along an azimuthal direction of Az165° (A-direction for the 2D analyses) and varying inclination angles. Shown are: (a) comparisons between TTFs and ETFs; and (b) goodness-of-fit metrics, $r$ and $m_{TF}$, for each comparison.

### 5.4 Effects of azimuthal angles on 3D GRA

To examine how azimuthal propagation direction influences the 3D site response for waves arriving with a fixed non-vertical inclination, GRAs were performed using a constant inclination angle of 15° while varying the azimuth (75°, 165°, 255°, and 345°). For each case, the simulated orthogonal EW and NS acceleration time histories at both the surface and downhole sensors were rotated into the corresponding input azimuth, and TTFs were computed from these azimuthally-resolved acceleration records to ensure consistency between the imposed propagation direction and the evaluated site response. Furthermore, in the present analyses, which involve multiple propagation directions and associated component rotations, a single azimuthally-resolved ETF is not appropriate for comparison. Instead, the comparisons presented herein use the overall lognormal median ETF, $LM_{ETF,EW\&NS}$ and its $\pm 1\sigma_{\ln ETF,EW\&NS}$ bounds, derived by combining all individual ETFs from both horizontal components (EW and NS), as described in Section 2 and shown in Figure 2d. This azimuth-independent reference ETF provides a consistent basis for comparison across all azimuthal cases.

The results are presented in Figures 12a and 12b. Figure 12b also includes a colormap of the $f_{0,H/V}$ values to illustrate spatial variations in bedrock depth, with higher $f_{0,H/V}$ values indicating shallower bedrock and lower values indicating deeper bedrock. This visualization aids interpretation of how the dipping bedrock structure in the southeastern portion of the site, together with the direction of horizontal wave propagation, may influence the site response. Figure 12a shows that azimuthal effects in 3D GRAs at DPDA are generally subtle compared with the influence of wave inclination. The $f_0$ and higher-mode peaks remain nearly unchanged across all azimuths, indicating that the vertically averaged impedance structure primarily controls the main resonance peaks. Some azimuth-dependent variability is observed in the trough between the ETF $f_0$ and $f_1$ peaks (approximately 1.5–2.5 Hz), where small secondary peaks appear; however, these differences are modest. Consistent with the qualitative observations, azimuthal effects are also minor relative to inclination effects in a quantitative sense, as shown in Figure 12b. All azimuthal cases except Az = 75° fall within a narrow range of Pearson correlation coefficients ($r = 0.82$–$0.83$), with the Az = 75° case yielding a slightly lower value ($r = 0.78$) due to a small shift in $f_0$. In terms of misfit, Az = 165° produces the lowest $m_{TF}$ value (1.28), compared with values ranging from 1.46 to 1.55 for the other azimuths. This improvement is largely attributable to the absence of a pronounced secondary peak for Az = 165°.

These results indicate that azimuthal direction does not appreciably affect the primary resonance frequencies or amplitudes at DPDA, and that azimuthal variability in wave incidence cannot explain the high apparent damping observed in the $f_0$ peak of the ETFs. Its influence is primarily confined to the structure of the trough between the $f_0$ and $f_1$ peaks, reflecting interactions among horizontal wave propagation, dipping bedrock geometry, and lateral $V_s$ gradients. These azimuth-related variations remain secondary and localized compared with the relatively stronger effects associated with wave inclination. Similar behavior was reported by Dawadi et al. (2026a) for the I-15 Downhole Array, where azimuthal variations produced only modest changes in site response predictions relative to inclination effects.

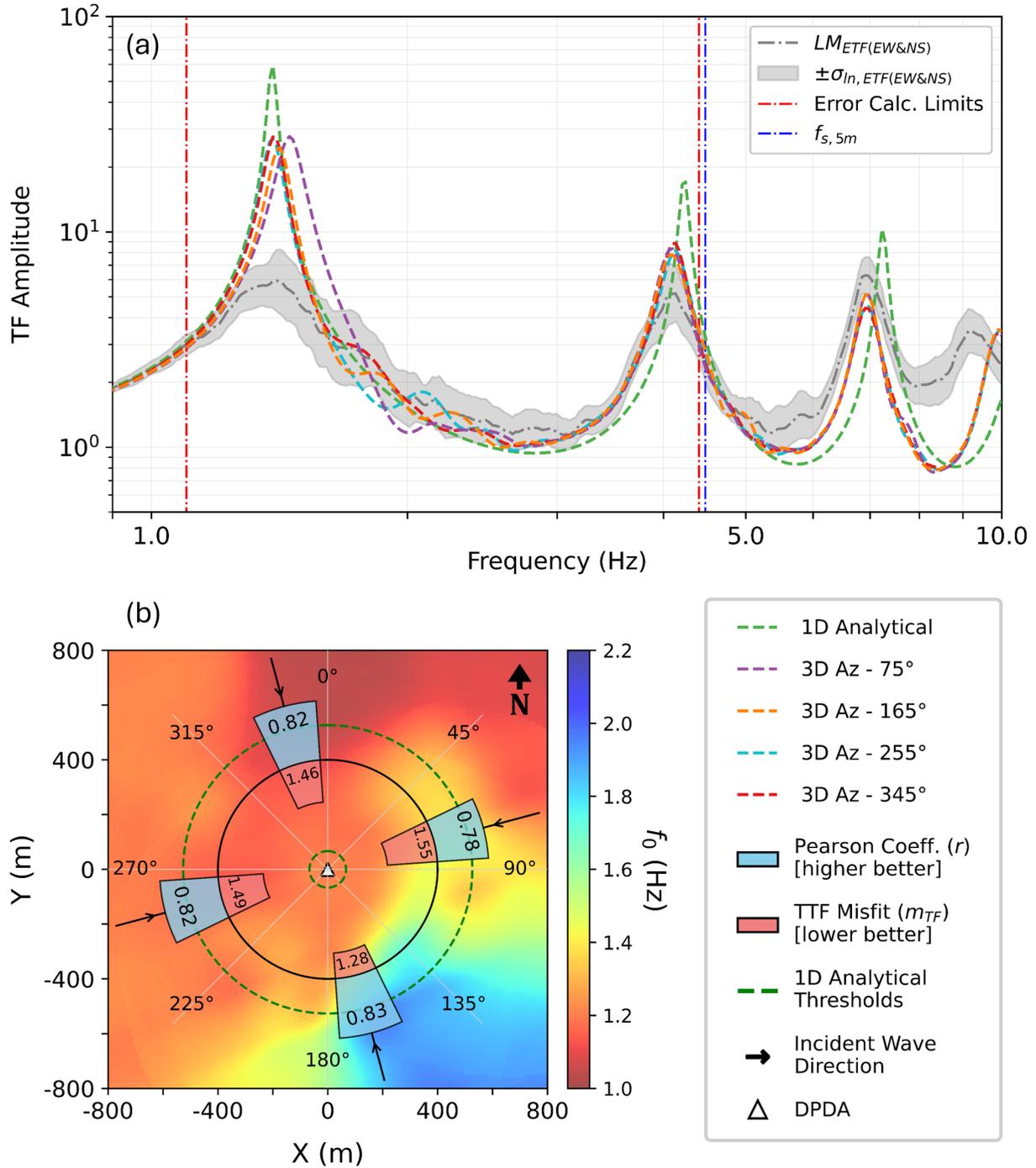

**Figure 12.** Site response prediction results from 3D GRAs at DPDA conducted using input motions from multiple azimuthal directions at an inclination angle of 15°. Shown are: (a) comparisons between TTFs and ETFs; and (b) goodness-of-fit metrics, $r$ and $m_{TF}$, for each comparison, plotted over the spatial distribution of the H/V-measurement-based fundamental frequencies ($f_{0,H/V}$). Higher $f_{0,H/V}$ values indicate shallower stiff layers or bedrock.

## 6. Conclusions

This study evaluated the effects of inclined and azimuthally varying incident waves on 2D and 3D GRAs at the DPDA site to assess whether modeling waves with non-vertical incidence can increase apparent damping and reduce the overestimation of TTF amplitudes relative to ETFs. The analyses were conducted using a large-scale pseudo-3D Vs model under linear-viscoelastic conditions in FLAC3D. The TTFs computed from the modeling were compared with ETFs calculated from small-strain ground motions recorded at multiple depths at the DPDA site. Two approaches were investigated for implementing inclined waves: the Input Lag Method (ILM) and the Inclined Domain Method (IDM). Although the IDM is theoretically appealing due to its compatibility with conventional free-field boundary formulations, it was found to be computationally prohibitive and susceptible to boundary-related artifacts when applied to kilometer-scale models. In contrast, the ILM consistently produced stable wavefields with minimal spurious reflections while avoiding domain rotation or expansion, making it more computationally efficient for inclined-wave simulations in large-scale 2D and 3D GRAs.

Results from the 2D GRAs indicate that the influence of incident wave inclination on site response is modest for small angles, with little change in amplification observed up to approximately 15°. Larger inclination angles produce noticeable reductions in peak amplitudes; however, these reductions are accompanied by systematic shifts in resonant frequencies that are not observed in the ETFs, suggesting that such steeply inclined incidence is unlikely to be representative of the DPDA ground motions used in this study. The 3D GRAs exhibit similar behavior, with comparable sensitivity to inclination and no fundamental change in the response trends relative to the 2D analyses. In both dimensions, modest inclination angles yield only limited improvement, whereas steep inclinations degrade agreement with the observed site response. Thus, it can be concluded that the recorded ground motions do not contain significant apparent attenuation from destructive interference associated with inclined waves, and that other factors must be influencing the poor amplitude agreement between TTFs and ETFs at the DPDA. These results also indicate that much of the benefit of multi-dimensional GRAs relative to 1D analyses are already captured within the 2D framework at DPDA, and that full 3D modeling does not provide a substantial additional improvement.

Analysis of azimuthally varying wave incidence in 3D indicates that horizontal propagation direction exerts only a secondary influence on site response at DPDA. Variations in input azimuth primarily affect the structure of the TTF troughs and the presence of minor secondary peaks, while the $f_0$ and $f_1$ peaks remain largely unchanged. Differences among azimuthal cases are modest relative to those associated with wave inclination, leading to minimal variation in site response across azimuthal input wave directions. Taken together, these results demonstrate that neither inclined-wave incidence nor azimuthal input variability, even when modeled in 3D, can fully account for the low amplification of ETF peaks observed at DPDA.

These findings highlight the need to incorporate additional attenuation mechanisms to reproduce the apparent damping observed at DPDA. Potential modeling approaches include deliberate amplification of intrinsic, laboratory-based damping to better represent apparent damping effects (Afshari and Stewart, 2019; Tao and Rathje, 2019), the use of alternative numerical damping formulations (Dawadi et al. 2026c), and explicit modeling of smaller-scale heterogeneities not captured by pseudo-3D $V_s$ models. Future work should, therefore, focus on modeling only slightly inclined-waves while simultaneously integrating modified damping formulations and more comprehensive representations of subsurface complexity.

**Acknowledgements**

The authors would like to thank the following individuals for their assistance with data acquisition and/or processing at the DPDA site over the years: Dr. Mohamad M. Hallal, Dr. Joseph Vantassel, and Dr. Michael Yust.

**CRediT authorship contribution statement**

Nishkarsha Dawadi: Investigation, Conceptualization, Data curation, Methodology, Software, Visualization, Validation, Formal Analysis, Writing – original draft, Writing – review & editing. Brady R. Cox: Investigation, Conceptualization, Funding acquisition, Resources, Project administration, Methodology, Supervision, Validation, Writing – review & editing

**Conflict of interest disclosure**

The author(s) can declare no potential conflicts of interest with respect to the research, authorship, and/or publication of this article.

**Research Funding**

This research was financially supported by Pacific Gas & Electric (PG&E). However, the opinions, findings, conclusions, and recommendations expressed in this paper are solely those of the authors and do not necessarily reflect the views or policies of PG&E.

**Data availability statement**

The raw site characterization data used to develop the pseudo-3D Vs models in this study are available in the DesignSafe Data Depot (Hallal et al., 2025).